%% file: report.tex
\documentclass[11pt]{article}

\usepackage{graphicx}
\usepackage{epsfig,amsmath,amssymb,amstext,amsthm,color}
\usepackage{subfigure}
\usepackage{floatflt}
\usepackage{verbatim}

\usepackage{url}

\urldef{\mailsa}\path|{krjampan, alubiw}@uwaterloo.ca|

{\bfseries}{\itshape}
\newtheorem{cor}{Corollory}{\bfseries}{\itshape}
{\bfseries}{\itshape}
{\bfseries}{\itshape}
{\bfseries}{\itshape}
\newtheorem{lemma}{Lemma}{\bfseries}{\itshape}
{\bfseries}{\itshape}

\title{Simultaneous Interval graphs}
\author{Krishnam Raju Jampani
\thanks{David R. Cheriton School of Computer Science,
University of Waterloo, Email:{\tt krjampan@uwaterloo.ca}}
\and Anna Lubiw \thanks{David R. Cheriton School of Computer Science,
University of Waterloo, Email:{\tt alubiw@uwaterloo.ca}}}
\date{}
\begin{document}
\pagestyle{plain}
\pagenumbering{arabic}
\maketitle

\newcommand{\mr}{\ensuremath{\mathbf{\eta}}}
\newcommand{\len}{\ensuremath{t}}
\newcommand{\minor}{\ensuremath{\preceq}}
\newcommand{\lrangle}[1]{\ensuremath{\langle #1 \rangle }}
\newcommand{\floor}[1]{\ensuremath{\lfloor #1 \rfloor }}
\newcommand{\ceil}[1]{\ensuremath{\lceil #1 \rceil }}
\newcommand{\cart}{\ensuremath{\,\Box\,}}
\newcommand{\brac}[1]{\ensuremath{\lbrace #1 \rbrace }}
\newcommand{\ignore}[1]{}

\begin{abstract}
 In a recent paper, we introduced the simultaneous representation problem (defined for any graph class
$\cal C$) and studied the problem for chordal, comparability and permutation graphs. For interval graphs,
the problem is defined as follows. Two interval graphs $G_1$ and $G_2$, sharing some vertices $I$ (and the
corresponding induced edges), are said to be
``simultaneous interval graphs'' if there exist interval representations $R_1$ and $R_2$ of $G_1$ and
$G_2$, such that any vertex of $I$ is mapped to the same interval in both $R_1$ and $R_2$. Equivalently,
$G_1$ and $G_2$ are simultaneous interval graphs if there exist edges $E'$ between $G_1-I$ and $G_2-I$
such that $G_1 \cup G_2 \cup E'$ is an interval graph. 

  Simultaneous representation problems are related to simultaneous planar embeddings, and have applications
in any situation where it is desirable to consistently represent two related graphs, for example: interval
graphs capturing overlaps of DNA fragments of two similar organisms; or graphs connected in time, where one
is an updated version of the other.

     In this paper we give an $O(n^2{\log n})$ time algorithm for recognizing  simultaneous interval graphs,
where $n = |G_1 \cup G_2|$. This result complements the polynomial time algorithms for recognizing probe
interval graphs and provides an efficient algorithm for the interval graph sandwich problem for the
special case where the set of optional edges induce a complete bipartite graph. \\

\noindent{\bf Keywords: Simultaneous Graphs, Interval Graphs, Graph Sandwich Problem, Probe Graphs, PQ-trees}
\end{abstract}

\input{main.tex}

\end{document}

%% file: main.tex
\section{Introduction}
\begin{figure}[h!]
\begin{center}
\scalebox{0.6}{\includegraphics{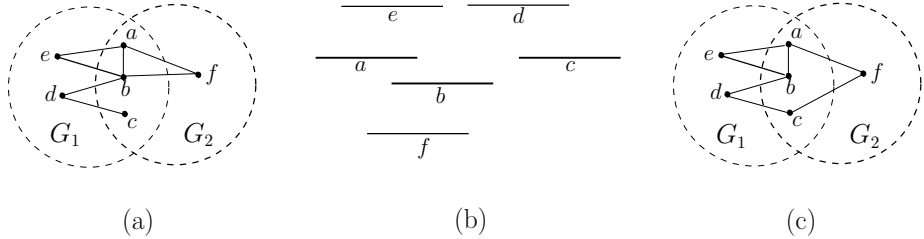}}
\end{center}
\caption{Graphs in (a) are simultaneous interval graphs as shown by the representations in (b). Graphs
in (c) are not simultaneous interval graphs.}
\label{FIG:intro_interval}
\end{figure}

 Let $\cal C$ be any intersection graph class (such as interval graphs or chordal graphs) and let
 $G_1$ and $G_2$ be two graphs in $\cal C$, sharing some vertices $I$ and the edges induced by $I$.
 $G_1$ and $G_2$ are said to be {\em simultaneously representable $\cal C$ graphs} or
 {\em simultaneous $\cal C$ graphs} if there exist intersection representations $R_1$ and $R_2$
 of $G_1$ and $G_2$ such that any vertex of $I$ is represented by the same object in both $R_1$
 and $R_2$. The {\em simultaneous representation problem} for class $\cal C$ asks whether $G_1$ and
$G_2$ are simultaneous $\cal C$ graphs. For example, Figures \ref{FIG:intro_interval}(a) and
\ref{FIG:intro_interval}(b) show two simultaneous interval graphs and their interval representations
with the property that vertices common to both graphs are assigned to the same interval. Figure
\ref{FIG:intro_interval}(c) shows two interval graphs that are not simultaneous interval graphs.

  Simultaneous representation problems were introduced by us in a recent paper~\cite{JL2} and have
application in any situation where two related graphs should be represented consistently. A main
instance is for temporal relationships, where an old graph and a new graph share some common parts.
Pairs of related graphs also arise in many other situations, e.g: two social networks that share
some members; overlap graphs of DNA fragments of two similar organisms, etc. 

  Simultaneous representations are related to simultaneous planar embeddings: two
graphs that share some vertices and edges (not necessarily induced) have a {\em simultaneous geometric
embedding}~\cite{BCDE} if they have planar straight-line drawings in which the common vertices are
represented by common points. Thus edges may cross, but only if they are in different graphs. Deciding
if two graphs have a simultaneous geometric embedding is NP-Hard~\cite{BGJPSS}.
 
In~\cite{JL2}, we showed that the simultaneous representation problem can be solved efficiently for chordal,
comparability and permutation graphs. We also showed that for any intersection class $\cal C$, the simultaneous
representation problem for $G_1$ and $G_2$ is equivalent to the following problem: Do there
exist edges $E'$ between $G_1-I$ and $G_2-I$ so that the augmented graph $G_1 \cup G_2 \cup E'$ belongs to class
$\cal C$.

  The {\em graph sandwich problem}~\cite{GKS} is a more general augmentation problem defined for any graph class $\cal C$: given graphs
$H_1 = (V, E_1)$ and $H_2= (V, E_2)$, is there a set $E$ of edges with $E_1 \subseteq E \subseteq E_2$ so that the graph $G = (V, E)$
belongs to class $\cal C$. This problem has a wealth of applications but is NP-complete for interval graphs, comparability graphs, and
permutation graphs~\cite{GKS}.
 
  The simultaneous representation problem (for class $\cal C$) is the special case of the graph
sandwich problem (for $\cal C$) where $E_2 - E_1$ forms a complete bipartite subgraph. A related
special case where $E_2 - E_1$ forms a clique is the problem of recognizing {\it probe graphs}:
a graph $G$ with a specified independent set $N$ is a {\it probe graph} for class $\cal C$ if
there exist edges $E' \subseteq N \times N$ so that the augmented graph $G \cup E'$ belongs to
class $\cal C$.

   Probe graphs have several applications~\cite{McMWZ,GL} and have received much attention recently.
The first polynomial-time algorithm for recognizing probe interval graphs was due to Johnson and
Spinrad~\cite{JS}. They used a variant of PQ-trees and achieved a run-time of $O(n^2)$. Techniques
from modular decomposition provided more speed up~\cite{McCS}, but the most recent algorithm by
McConnell and Nussbaum~\cite{McCN} reverts to PQ-trees and achieves linear time.

  We note that there has also been work~\cite{Ta} on a concept of simultaneous intersection called ``polysemy'' where
two graphs are represented as intersections of sets and their complements.

In this paper, we give an $O(n^2 {\log n})$ algorithm for solving the simultaneous representation problem
for interval graphs. We use PQ-trees, which were developed by Booth and Lueker for the original linear time
interval graph recognition algorithm. They used a PQ-tree to capture the orderings of the maximal cliques
of the graph (see~\cite{Golumbic} for an introduction to interval graphs and PQ-trees). 

In the probe interval recognition problem, there is a single PQ-tree (of the graph induced by the probes) and a set
of constraints imposed by the non-probes. However in our situation we have two PQ-trees, one for each graph, that we want 
to re-order to ``match'' on the common vertex set $I$. 
We begin by ``reducing'' each PQ-tree to contain only vertices from $I$. This results in PQ-trees that store non-maximal
cliques, and our task is to modify each PQ-tree by inserting non-maximal cliques from the other tree while re-ordering the
trees to make them the same.

\section{Reduction to PQ-trees}
  In this section we transform the simultaneous interval graph problem to a problem about ``compatibility'' of two PQ-trees arising from
the two graphs.
 
  Recall that an interval graph is defined to be the intersection graph of
intervals on the real line.   For any point on the line, the intervals containing that point form a clique in the graph.  This leads to the fundamental one-to-one correspondence between the interval representations of an interval graph and its {\em clique orderings}, defined as follows:  A {\em clique ordering} of $G$ is a sequence of (possibly empty) cliques ${\cal S} = Q_1, Q_2, \cdots, Q_l$ that contains all the maximal cliques of $G$ and has the property that for each vertex $v$, the cliques in $\cal S$ that contain $v$ appear consecutively. Note that we allow cliques to be empty.

The standard interval graph recognition algorithm attempts to find a clique order of the maximal cliques of a graph by making the maximal cliques into leaves of a PQ-tree, and imposing PQ-tree constraints to ensure that the cliques containing each vertex $v$ appear consecutively. This structure is called {\em the} PQ-tree of the graph. Note that the children of a P-node may be reordered
arbitrarily and the children of a Q-node may only be reversed. We consider a node with 2 children to be a Q-node. In the figures, we use a
circle to denote a P-node and a rectangle to denote a Q-node. A {\em leaf-order} of a PQ-tree is the order in which its
leaves are visited in an in-order traversal of the tree, after children of P and Q-nodes are re-ordered as just
described. 

Note that ignoring non-maximal cliques is fine for recognizing interval graphs; for our purposes, however, we want to consider clique orders and PQ-trees that may include non-maximal cliques.  We say that a PQ-tree whose leaves correspond to cliques of a graph is {\em valid} if for each of its leaf orderings and for each vertex $v$, the cliques containing $v$ appear consecutively.
 
 Let ${\cal S} = Q_1, Q_2, \cdots, Q_l$ be a clique ordering of interval graph $G$ and let the maximal cliques of $G$ be
$Q_{i_1}, Q_{i_2}, \cdots, Q_{i_m}$ (appearing in positions $i_1 < i_2 < \cdots < i_m$ respectively). Note that all the cliques in $\cal S$
between $Q_{i_j}$ and $Q_{i_{j+1}}$  contain $B = Q_{i_j} \cap Q_{i_{j+1}}$. We say that
$B$ is the {\em boundary clique} or {\em boundary} between $Q_{i_j}$ and $Q_{i_{j+1}}$.
Note that $B$ may not necessarily be present in $\cal S$.
The sequence of cliques between $Q_{i_j}$ and $Q_{i_{j+1}}$ that are  subsets
of $Q_{i_j}$ is said to be the {\em right tail} of $Q_{i_j}$. The {\em left tail} of $Q_{i_{j+1}}$ is defined analogously.
Observe that the left tail of a clique forms an increasing sequence and the right tail forms a decreasing sequence (w.r.t set inclusion).
Also note that all the cliques that precede $Q_{i_1}$ are subsets of $Q_{i_1}$ and this sequence is called the left tail of $Q_{i_1}$ and
all the cliques that succeed $Q_{i_m}$ are subsets of $Q_{i_m}$ and this sequence is called the right tail of $Q_{i_m}$. Thus any clique ordering of $G$
consists of a sequence of maximal cliques, with each maximal clique containing a (possibly empty) left and right tail of subcliques.

  Let $Q_0$ and
$Q_{l+1}$ be defined to be empty sets. An insertion of clique $Q'$ between $Q_i$ and $Q_{i+1}$ (for some $i \in \brac{0,\cdots, l}$)
is said to be a {\em subclique insertion} if $Q' \supseteq Q_i \cap Q_{i+1}$ and either $Q' \subseteq Q_i$ or $Q' \subseteq Q_{i+1}$.
It is clear that after a subclique insertion the resulting sequence is still a clique ordering of $G$.
A clique ordering ${\cal S'}$ is an {\it extension} of ${\cal S}$ if ${\cal S}'$ can be obtained from $\cal S$ by subclique
insertions. We also say that ${\cal S}$ extends to ${\cal S}'$. Furthermore, we say that a clique ordering is {\em generated} by a PQ-tree,
if it can be obtained from a leaf order of the PQ-tree with subclique insertions. The above definitions yield the following Lemma.

\begin{lemma}
\label{clique-ordering}
A sequence of cliques $\cal S$ is a clique ordering of $G$ if and only if $\cal S$ can be generated from the PQ-tree of $G$.
\end{lemma}

Let $G_1$ and $G_2$ be two interval graphs sharing a vertex set $I$ (i.e.~$I= V(G_1) \cap V(G_2)$) and its induced edges.
Note that $G_1[I]$ is isomorphic to $G_2[I]$. A clique ordering of $G_1[I]$ is said to be an  $I$-ordering.

The $I$-{\em restricted} PQ-tree of $G_j$ is defined to be the tree obtained from the PQ-tree of $G_j$ by replacing each clique $Q$ 
(a leaf of the PQ-tree) with the clique $Q \cap I$. Thus there is a one-to-one correspondence between the two PQ-trees, and 
the leaves of the $I$-restricted PQ-tree are cliques of $G_1[I]$.
 
 Let ${\cal I} = X_1, X_2, \cdots, X_l$ be an $I$-ordering. ${\cal I}$ is said to be $G_j$-{\em expandable} if there exists a
clique ordering ${\cal O} = Q_1, Q_2, \cdots, Q_l$ of $G_j$ such that $X_i \subseteq Q_i$ for $i \in \brac{1,\cdots,l}$.
Further, we say that $\cal I$ expands to $\cal O$.
By the definition of clique-ordering it follows that, if ${\cal I}$ is $G_j$-expandable then it remains $G_j$-expandable
after a subclique insertion (i.e.~any extension of $\cal I$ is also $G_j$-expandable). We first observe the following.

\begin{lemma}
\label{expandable-ordering}
The set of $G_j$-expandable $I$-orderings is same as the set of orderings that can be generated from the $I$-restricted PQ-tree of $G_j$.
\end{lemma}
\begin{proof}
 Let $T$ be the PQ-tree of $G_j$ and $T'$ be the $I$-restricted PQ-tree of $G_j$.

 Let $\cal I$ be a $G_j$-expandable $I$-ordering of $G_j$. Then there exists a clique ordering $\cal O$ of $G_j$ such that $\cal I$ expands
to $\cal O$. But by Lemma \ref{clique-ordering}, $\cal O$ can be generated from $T$ (from a leaf order with subclique insertions).
This in turn implies that $\cal I$ can be generated from $T'$ (from the corresponding leaf order with the corresponding subclique
insertions).

 Now for the other direction, let ${\cal I}' = X_1, \cdots, X_l$ be any leaf order of $T'$. Then there exists a corresponding leaf
order ${\cal O}' = Q_1, \cdots, Q_l$ of $T$ such that $X_i \subseteq Q_i$ for $i \in \brac{1, \cdots, l}$. This implies that ${\cal I}'$
is a $G_j$-expandable $I$-ordering. Finally, observe that if $I''$ is generated from $I'$ by subclique insertions than $I''$ is also a
$G_j$-expandable $I$-ordering. Thus the Lemma holds.
\end{proof}

 Two $I$-orderings ${\cal I}_1$ and ${\cal I}_2$ are said to be {\em compatible} if both ${\cal I}_1$ and ${\cal I}_2$
(separately) extend to a common $I$-ordering $\cal I$. For e.g.~the ordering $\brac{1},\brac{1,2},\brac{1,2,3,4}$ is compatible 
with the ordering $\brac{1}, \brac{1,2,3}, \brac{1,2,3,4}$, as they both extend to the common ordering: 
$\brac{1}, \brac{1,2}, \brac{1,2,3}, \brac{1,2,3,4}$. Note that the compatibility relation is not transitive.
Two PQ-trees $T_1$ and $T_2$ are said to be {\em compatible} if there exist orderings ${\cal O}_1$ and ${\cal O}_2$
generated from $T_1$ and $T_2$ (respectively) such that ${\cal O}_1$ is compatible with ${\cal O}_2$. The following Lemma
is our main tool.

\begin{lemma}
\label{restrictedPQ}
$G_1$ and $G_2$ are simultaneous interval graphs if and only if the $I$-restricted PQ-tree of $G_1$ is compatible with the
$I$-restricted PQ-tree of $G_2$.
\end{lemma}
\begin{proof}
    By Lemma \ref{expandable-ordering}, it is enough to show that $G_1$ and $G_2$ are simultaneous interval graphs if and only if there
exists a $G_1$-expandable $I$-ordering ${\cal I}_1$ and a $G_2$-expandable $I$-ordering ${\cal I}_2$ such that ${\cal I}_1$ is
compatible with ${\cal I}_2$. We now show this claim.

 Let ${\cal I}_1$ and ${\cal I}_2$ be as defined in the hypothesis. Since ${\cal I}_1$ and ${\cal I}_2$ are compatible, they can
be extended to a common $I$-ordering $\cal I$. Let $\cal I$
expand to clique orderings ${\cal O}_1$ and ${\cal O}_2$ in $G_1$ and $G_2$ respectively. Since each vertex of $I$ appears
in the same positions in both ${\cal O}_1$ and ${\cal O}_2$, it is possible to obtain interval representations $R_1$ and $R_2$ of $G_1$
and $G_2$ (from ${\cal O}_1$ and ${\cal O}_2$ respectively) such that each vertex in $I$ has the same end points in both $R_1$
and $R_2$. This implies that $G_1$ and $G_2$ are simultaneous interval graphs.

 For the other direction, let $G_1$ and $G_2$ be simultaneous interval graphs. Then there exists an augmenting set of edges
$A' \subseteq V_1 - I \times V_2 - I$ such that $G= G_1 \cup G_2 \cup A'$ is an interval graph. Let ${\cal O} = Q_1, Q_2, \cdots, Q_l$
be a clique-ordering of $G$. For each $i \in \brac{1,\cdots,l}$ and $j \in \brac{1,2}$, by restricting $Q_i$ to $V_j$ (i.e.~replacing
$Q_i$ with $Q_i \cap V_j$), we obtain a clique ordering ${\cal O}_j$ of $G_j$. Now for $j \in {1,2}$, let ${\cal I}_j$ be the
$I$-ordering obtained from ${\cal O}_j$ by restricting each clique in ${\cal O}_j$ to $I$. It follows that ${\cal I}_1$ is a
$G_1$-expandable $I$-ordering and ${\cal I}_2$ is a $G_2$-expandable $I$-ordering. Further ${\cal I}_1 = {\cal I}_2$ and hence
${\cal I}_1$ and ${\cal I}_2$ are compatible.
\end{proof}

Our algorithm will decide if the $I$-restricted PQ-tree of $G_1$ is compatible with the $I$-restricted
PQ-tree of $G_2$. We first show how the $I$-restricted PQ-trees can be simplified in several ways.
Two $I$-orderings ${\cal I}_1$ and ${\cal I}_2$ are said to be {\em equivalent} if for any $I$-ordering ${\cal I}'$, ${\cal I}_1$ and ${\cal I}'$
are compatible if and only if ${\cal I}_2$ and ${\cal I}'$ are compatible. Note that this is an equivalence relation. The Lemma below follows
directly from the definitions of equivalent orderings and subclique insertions.

\begin{lemma}
\label{duplicates}
Let ${\cal I} = X_1, X_2, \cdots, X_l$ be an $I$-ordering in which $X_i = X_{i+1}$ for some $i \in {1,\cdots, l-1}$. Let $\cal I'$ be
the $I$-ordering obtained from $\cal I$ by deleting $X_{i+1}$. Then $\cal I$ is equivalent to $\cal I'$.
\end{lemma}

Further, because equivalence is transitive, Lemma \ref{duplicates} implies that an $I$-ordering $\cal I$ is equivalent to the $I$-ordering
$\cal I'$ in which all consecutive duplicates are eliminated. This allows us to simplify the $I$-restricted PQ-tree of $G_j$.
Let $T$ be the $I$-restricted PQ-tree of $G_j$. We obtain a PQ-tree $T'$ from $T$ as follows. \\

\noindent
1. Initialize $T' = T$. \\
\noindent
2. As long as there is a non-leaf node $n$ in $T'$ such that all the descendants of $n$ are the same, i.e.~they are all duplicates of a
single clique $X$, replace $n$ and the subtree rooted at $n$ by a leaf node representing $X$. \\
\noindent
3. As long as there is a (non-leaf) Q-node $n$ in $T'$ with two consecutive child nodes $n_a$ and $n_b$ (among others) such that all the
descendants of $n_a$ and $n_b$ are the same i.e.~they are all duplicates of a single clique $X$, replace $n_a$, $n_b$ and the subtrees
rooted at these vertices by a single leaf node representing the clique $X$. \\

 Note that the resulting $T'$ is unique. We call $T'$ the $I$-reduced PQ-tree of $G_j$.

\begin{lemma}
\label{reducedPQ}
$G_1$ and $G_2$ are simultaneous interval graphs if and only if the $I$-reduced PQ-tree of $G_1$ is compatible with the $I$-reduced PQ-tree
of $G_2$.
\end{lemma}
\begin{proof}
For $j \in \brac{1,2}$, let $T_j$ and $T'_j$ be the $I$-restricted and $I$-reduced PQ-trees of $G_j$ respectively. Let $\cal I$ be any
$I$-ordering. Observe that by Lemma \ref{duplicates}, $\cal I$ is compatible with a leaf ordering of $T_j$ if and only if $\cal I$
is compatible with a leaf ordering of $T'_j$. Thus the conclusion follows from Lemma \ref{restrictedPQ}.
\end{proof}

\section{Labeling and Further Simplification}

 In section 2, we transformed the simultaneous interval graph problem to a problem of testing compatibility of two $I$-reduced PQ-trees where
$I$ is the common vertex set of the two graphs. These PQ-trees may have nodes that correspond to non-maximal cliques in $I$. In this section we prove
some basic properties of such $I$-reduced PQ-trees, and use them to further simplify each tree.

 Let $\cal T$ be the $I$-reduced PQ-tree of $G_j$.
Recall that each leaf $l$ of $\cal T$ corresponds to a clique $X$ in $G_j[I]$.
If $X$ is maximal in $I$, then $X$ is said to be a {\em max-clique} and $l$ is said to be a {\em max-clique node}, otherwise
$X$ is said to be a {\em subclique} and $l$ is said to be a {\em subclique node}. When the association is clear from the context,
we will sometimes refer to a leaf $l$ and its corresponding clique $X$ interchangeably, or interchange the terms ``max-clique'' and
``max-clique node'' [resp. subclique and subclique node]. A node of $\cal T$ is said to be an {\em essential node} if it is a non-leaf node
or if it is a leaf node representing a max-clique.

 Given a node $n$ of $\cal T$, the {\it descendant cliques} of $n$ are the set of cliques that correspond to the leaf-descendants
of $n$. Because our algorithm operates by inserting subcliques from one tree into the other, we must take care to preserve the validity
of a PQ-tree. For this we need to re-structure the tree when we do subclique insertions. The required restructuring will be determined
based on the label $U(n)$ that we assign to each node $n$ as follows. \\

\noindent
 $U(n)$ or the {\it Universal set} of $n$ is defined as the set of vertices $v$ such that $v$ appears in all descendant cliques of $n$. \\

Note that for a leaf node $l$ representing a clique $X$, $U(l) = X$ by definition. Also note that along any path
up the tree, the universal sets decrease. The following Lemma gives some useful properties of the $I$-reduced 
PQ-tree.

\begin{lemma}
\label{preliminaries}
Let $\cal T$ be the $I$-reduced PQ-tree of $G_j$. Let $n$ be a non-leaf node of $\cal T$ ($n$ is used in properties 2--6). Then we have: \\
\noindent
{\bf 0.} Let $l_1$ and $l_2$ be two distinct leaf nodes of $\cal T$, containing a vertex $t \in I$. Let $y$ be the least common ancestor of $l_1$ and
$l_2$. Then: (a) If $y$ is a P-node then all of its descendant cliques contain $t$. (b) If $y$ is a Q-node then $t$ is contained in all 
the descendant cliques of all children of $y$ between (and including) the child of $y$ that is the ancestor of $l_1$ and the child that is the
ancestor of $l_2$. \\
\noindent
{\bf 1.} Each max-clique is represented by a unique node of $\cal T$. \\
\noindent
{\bf 2.} A vertex $u$ is in $U(n)$ if and only if for every child $n_1$ of $n$, $u \in U(n_1)$. \\
\noindent
{\bf 3.} $n$ contains a max-clique as a descendant. \\
\noindent
{\bf 4.} If $n$ is a P-node, then for any two child nodes $n_1$ and $n_2$ of $n$, we have $U(n) = U(n_1) \cap U(n_2)$. \\
\noindent
{\bf 5.} If $n$ is a P-node, then any child of $n$ that is a subclique node represents the clique $U(n)$. \\
\noindent
{\bf 6.} If $n$ is a Q-node and $n_1$ and $n_2$ are the first and last child nodes of $n$ then $U(n) = U(n_1) \cap U(n_2)$.
\end{lemma}
\begin{proof}
(0) Observe that in any leaf ordering of $\cal T$, all the nodes that appear between $l_1$ and $l_2$ must also contain the vertex $t$,
otherwise $\cal T$ would be invalid. Now let $l_3$ be a leaf descendant of $y$, that doesn't contain $t$.

    If $y$ is a P-node, then we can reorder the children of $y$ in such a way that in the leaf-ordering of the resulting tree $l_3$ appears
between $l_1$ and $l_2$. But this contradicts the validity of $\cal T$. This proves (a). Similarly, if $y$ is a Q-node, then $l_3$ cannot be
equal to $l_1$ or $l_2$ or any node between them. Thus (b) also holds. \\

(1) Note that by definition of $I$-reduced PQ-tree of $G_j$, each max-clique must be present in $\cal T$. Now assume for the sake of contradiction
that a max-clique $X$ is represented by two leaf nodes, say $l_1$ and $l_2$. Let $y$ be the least common ancestor of $l_1$ and $l_2$. Let $c_1$
and $c_2$ be the child nodes of $y$ that contain $n_1$ and $n_2$ (respectively) as descendants. Now by (0), if $y$ is a P-node then all of its
descendant cliques must contain all the vertices of $X$. But as $X$ is maximal, all these cliques must be precisely $X$. However this is not
possible, as we would have replaced $y$ with a leaf node representing $X$ in the construction of $\cal T$. Similarly, if $y$ is a Q-node then the descendant cliques of $c_1$, $c_2$ and all the nodes between them must represent the max-clique $X$. But then we would have replaced these nodes with
with a leaf node representing $X$ in the construction of $\cal T$. This proves (1). \\

(2) If $u \in U(n)$, then all the descendant cliques of $n$ contain $u$. This implies that for any child $n_1$ of $n$, all the
descendant cliques of $n_1$ contain $u$. Hence $u \in U(n_1)$. On the other hand, if each child $n_1$ of $n$ contains a vertex $u$ in its
universal set, then $u$ is present in all the descendant cliques of $n$ and thus $u \in U(n)$. \\

(3) Note that if each descendant clique of $n$ contains precisely $U(n)$ (and no other vertex), then we would have replaced the subtree 
rooted at $n$ with a leaf node corresponding to the clique $U(n)$, when constructing $\cal T$. Thus there exists a clique $Q_2$ that is
a descendant of $n$, such that $Q_2-U(n)$ is non-empty. If $Q_2$ is a max-clique then we are done. Otherwise let $t \in Q_2-U(n)$ and 
let $Q_1$ be a max-clique containing $t$. Suppose $Q_1$ is a not a descendant of $n_1$. Applying (0) on $Q_1$ and $Q_2$, we infer that 
irrespective of whether $n$ is a P-node or a Q-node, all the descendant cliques of $n$ must contain $t$. But then $t \in U(n)$, a 
contradiction. Thus $Q_1$ is a descendant of $n_1$. \\

(4) By (2) we observe that $U(n) \subseteq U(n_1) \cap U(n_2)$. Thus it is enough to show that $U(n_1) \cap U(n_2) \subseteq U(n)$.
Let $u \in U(n_1) \cap U(n_2)$, then $u$ is present in all the descendant cliques of $n_1$ and $n_2$. By (0), $u$ must be present
in all the descendant cliques of $n$ and hence $u \in U(n)$. Therefore $U(n_1) \cap U(n_2) \subseteq U(n)$. \\

(5) Consider any child $n_1$ of $n$. Suppose $n_1$ is a leaf-node and is not a max-clique. It is enough to show that $n_1$ represents
the clique $U(n)$ i.e.~$U(n_1) = U(n)$. Suppose not. Then there exists a vertex $t \in U(n_1)-U(n)$.  Let $Q_1$ be a max-clique
containing $t$. Note that the common ancestor of $Q_1$ and $n_1$ is either $n$ or an ancestor of $n$. Applying (0) on $Q_1$ and $n_1$,
we infer that all the descendant cliques of $n$ must contain $t$. But then $t \in U(n)$, a contradiction. \\

(6) This follows from (2) and (0).
\end{proof}

Let $\cal T$ be the $I$-reduced PQ-tree of $G_j$.  Recall that an essential node is a non-leaf node or
a leaf node representing a maximal clique. Equivalently (by Lemma \ref{preliminaries}.3), an essential node is a node which contains
a max-clique as a descendant. The following Lemma shows that in some situations we can obtain an equivalent tree by
deleting subclique child nodes of a P-node $n$. Recall that by Lemma \ref{preliminaries}.5, such subclique nodes represent the clique $U(n)$.
 
\begin{lemma}
\label{furthersimplification}
 Let $\cal T$ be the $I$-reduced PQ-tree of $G_j$ and $n$ be a P-node in $\cal T$. Then \\
\noindent
{\bf 1.} If $n$ has at least two essential child nodes, then $\cal T$ is equivalent to the tree ${\cal T}'$,
obtained from $\cal T$ by deleting all the subclique children of $n$. \\
\noindent
{\bf 2.} If $n$ has at least two subclique child nodes, then $\cal T$ is
equivalent to the tree ${\cal T'}$, obtained from $\cal T$ by deleting all except one of the subclique children of $n$.
\end{lemma}
\begin{proof}
 We give the proof of (1) below. The proof of (2) is very similar and hence omitted.

 Let ${\cal O}_1$ be any $I$-ordering. It is enough to show that there exists a leaf ordering $\cal O$ of $\cal T$ that is compatible with
${\cal O}_1$ if and only if there exists a leaf ordering ${\cal O}'$ of ${\cal T}'$ that is compatible with ${\cal O}_1$.

 Let $\cal O$ be any leaf ordering of $\cal T$, compatible with ${\cal O}_1$. Consider the ordering ${\cal O}'$ obtained from $\cal O$ by
deleting the cliques $U(n)$ that correspond to the child nodes of $n$ in $\cal T$. Clearly ${\cal O}'$ is a leaf ordering of ${\cal T}'$.
Further ${\cal O}'$ can be extended to $\cal O$ by adding copies of the cliques $U(n)$ at appropriate positions. Thus ${\cal O}'$ is
compatible with ${\cal O}_1$.

 Now for the other direction, let ${\cal O}'$ be a leaf order of ${\cal T}'$, compatible with ${\cal O}_1$ and let ${\cal O}'$ and
${\cal O}_1$ extend to a common ordering ${\cal O}_F$. From the hypothesis, we can assume that there exist two essential child nodes
$n_1$ and $n_2$ of $n$ in ${\cal T}'$ such that the clique descendants of $n_1$, appear immediately before the clique descendants of
$n_2$ in ${\cal O}'$. Also let $S(n_1)$ and $S(n_2)$ be the
two subsequences of ${\cal O}'$ containing the clique descendants of $n_1$ and $n_2$ respectively. Since $n_1$ and $n_2$ are essential
nodes, $S(n_1)$ and $S(n_2)$ each contain at least one max-clique. Let $Q_1$ be the last max-clique in $S(n_1)$ and $Q_2$ be the first
max-clique in $S(n_2)$. By Lemma
\ref{preliminaries}.0, $Q_1 \cap Q_2 = U(n_1) \cap U(n_2) = U(n)$. Since ${\cal O}'$ is compatible with ${\cal O}_1$, in each of the two
orderings ${\cal O}'$ and ${\cal O}_1$, $Q_2$ occurs after $Q_1$ and no other max-clique appears between them. Further the
same holds for ${\cal O}_F$ (as it is an extension of ${\cal O}'$). Let $k$ be the number of subclique children of $n$ (that represent
the clique $U(n)$). Then
obtain a leaf ordering $\cal O$ of $\cal T$, from ${\cal O}'$, by inserting $k$ copies of $U(n)$ between $S(n_1)$ and $S(n_2)$. Now extend
${\cal O}_F$ to ${\cal O}'_F$ by inserting $k$ copies of $U(n)$ between $Q_1$ and $Q_2$ (there is a unique way of adding a subclique
between two max-cliques). It is clear that ${\cal O}'_F$ is an extension of both $\cal O$ and ${\cal O}_1$. Therefore ${\cal O}$ is
compatible with ${\cal O}_1$. This proves (1).
\end{proof}

We will simplify $\cal T$ as much as possible by applying Lemma \ref{furthersimplification} and by converting nodes with two children
into Q-nodes. We call the end result a {\em simplified} $I$-reduced PQ-tree, but continue to use the term ``$I$-reduced PQ-tree'' to
refer to it. Note that the simplification process does not change the universal sets and
preserves the validity of the PQ-tree so Lemma \ref{reducedPQ} and all the properties given in Lemma \ref{preliminaries} still hold.
Because we consider nodes with 2 children as Q-nodes Lemma \ref{furthersimplification} implies:

\begin{cor}
\label{simplification}
In a [simplified] $I$-reduced PQ-tree, any P-node has at least 3 children, and all the children are essential nodes.
\end{cor}

\section{Algorithm}
   For $k \in \brac{1,2}$, let ${\cal T}_k$ be the [simplified] $I$-reduced PQ-tree of $G_k$. By Lemma \ref{reducedPQ}, testing whether $G_1$ and $G_2$
are simultaneous interval graphs is equivalent to testing whether ${\cal T}_1$ and ${\cal T}_2$ are compatible. We test this by modifying ${\cal T}_1$
and ${\cal T}_2$ (e.g.~inserting the sub-clique nodes from one tree into the other) so as to make them identical, without losing their compatibility.
The following is a high level overview of our approach for checking whether ${\cal T}_1$ and ${\cal T}_2$ are compatible.

 
   Our algorithm is iterative and tries to {\em match} essential nodes of ${\cal T}_1$ with essential nodes of ${\cal T}_2$ in a bottom-up
fashion. An essential node $n_1$ of ${\cal T}_1$ is matched with an essential node $n_2$ of ${\cal T}_2$ if and only if the subtrees rooted
at $n_1$ and $n_2$
are the same i.e.~their essential children are matched, their subclique children are the same and furthermore (in the case of Q-nodes) 
their child nodes appear in the same order. If $n_1$ is matched with $n_2$ then we consider $n_1$ and $n_2$ to be identical and use the same name
(say $n_1$) to refer to either of them. Initially, we match each max-clique node of ${\cal T}_1$ with the corresponding max-clique node
of ${\cal T}_2$. Note that every max-clique node appears uniquely in each tree by Lemma \ref{preliminaries}.1. A sub-clique node may
appear in only one tree in which case we must first insert it into the other tree. This is done when we consider the parent of the
subclique node. 

In each iteration, we either match an unmatched node $u$ of ${\cal T}_1$ to an unmatched node $v$ of ${\cal T}_2$ (which may
involve inserting subclique child nodes of $v$ as child nodes of $u$ and vice versa) or we {\em reduce}
either ${\cal T}_1$ or ${\cal T}_2$ without losing their compatibility relationship. {\em Reducing} a PQ-tree means restricting it to reduce the
number of leaf orderings. Finally, at the end of the algorithm either we have modified
${\cal T}_1$ and ${\cal T}_2$ to a ``common'' tree ${\cal T}_I$  that establishes
their compatibility or we conclude that ${\cal T}_1$ is not compatible with ${\cal T}_2$. The common tree ${\cal T}_I$ is said to be an 
{\em intersection tree}
(of ${\cal T}_1$ and ${\cal T}_2$) and has the property that any ordering generated by ${\cal T}_I$ can also be generated by ${\cal T}_1$ and
${\cal T}_2$. If ${\cal T}_1$ and ${\cal T}_2$ are compatible, there may be several intersection trees of ${\cal T}_1$ and ${\cal T}_2$, but
our algorithm finds only one of them.

We need the following additional notation for the rest of this paper. A sequence of subcliques ${\cal S} = X_1, X_2, \cdots, X_l$ is said
to satisfy the {\em subset property} if $X_i \subseteq X_{i+1}$ for $i \in \brac{1,\cdots,l-1}$. $\cal S$ is said to satisfy the
{\em superset property} if $X_i \supseteq X_{i+1}$ for each $i$. Note that $S$ satisfies the subset property if and only if
$\bar{\cal S} = X_l, \cdots, X_2, X_1$ satisfies the superset property.

 Let $d$ be an essential child node of a Q-node in ${\cal T}_k$.
We will overload the term ``tail'' (previously defined for a max clique in a clique ordering) and define the {\em tails} of $d$ as follows.
The left tail (resp. right tail) of $d$ is defined as the sequence of subcliques that appear as siblings of $d$, to the immediate left (resp. right)
of $d$, such that each subclique is a subset of $U(d)$.  Note that the left tail of $d$ should satisfy the subset property and
the right tail of $d$ should satisfy the superset property (otherwise ${\cal T}_k$ will not be valid). Also note that since the children of
a Q-node can be reversed in order, ``left''
and ``right'' are relative to the child ordering of the Q-node. We will be careful to use ``left tail'' and ``right tail'' in such a way that this
ambiguity does not matter.   Now suppose $d$ is a matched node. Then in order to match the parent of $d$ in ${\cal T}_1$ with the parent of $d$
in ${\cal T}_2$, our algorithm has to ``merge'' the tails of $d$.

  Let ${\cal L}_1$ and ${\cal L}_2$ be two subclique sequences that satisfy the subset property. Then ${\cal L}_1$ is said to
be {\em mergable} with ${\cal L}_2$ if the union of subcliques in ${\cal L}_1$ and ${\cal L}_2$ can be arranged into an ordering
${\cal L}'$ that satisfies the subset property. Analogously, if ${\cal L}_1$ and ${\cal L}_2$ satisfy the superset property, then
they are said to be mergable if the union of their subcliques can be arranged into an ordering ${\cal L}'$ that satisfies the
superset property. In both cases, ${\cal L}'$ is said to be the {\em merge} of ${\cal L}_1$ and ${\cal L}_2$ and is denoted by
${\cal L}_1 + {\cal L}_2$.

  A {\em maximal matched node} is a node that is matched but whose parent is not matched. For an unmatched essential node $x$, the 
{\em MM-descendants} of $x$, denoted by $MMD(x)$ are its descendants that are maximal matched nodes. If $x$ is matched then we define
$MMD(x)$ to be the singleton set containing $x$. Note that the MM-descendants of an essential node is non-empty (since every essential 
node has a max-clique descendant).

  Our algorithm matches nodes from the leaves up, and starts by matching the leaves that are max-cliques. As the next node $n_1$ that
we try to match, we want an unmatched node whose essential children are already matched.
To help us choose between ${\cal T}_1$ and ${\cal T}_2$, and also to break ties, we prefer a node with larger $U$ set. Then,
as a candidate to match $n_1$ to, we want an unmatched node in the other tree that has some matched children in common with $n_1$. With
this intuition in mind, our specific rule is as follows.

 Among all the unmatched essential nodes of ${\cal T}_1$ union ${\cal T}_2$ choose $n_1$ with maximal $U(n_1)$, minimal $MMD(n_1)$, and
maximal depth, in that preference order. Assume without loss of generality that $n_1 \in {\cal T}_1$. Select an unmatched node $n_2$ 
from ${\cal T}_2$ with maximal $U(n_2)$, minimal $MMD(n_2)$ and maximal depth (in that order) satisfying the property that 
$MMD(n_1) \cap MMD(n_2) \not= \emptyset$. The following Lemma captures
certain properties of $n_1$ and $n_2$, including why these rules match our intuitive justification.

\begin{lemma}
\label{properties}
For $n_1$ and $n_2$ chosen as described above,
let $M_1 = MMD(n_1)$, $M_2 = MMD(n_2)$ and $X = M_1 \cap M_2$. Also let $C_1$ and $C_2$
be the essential child nodes of $n_1$ and $n_2$ respectively. Then we have: \\
\noindent
{\bf 1.} $M_1 = C_1$ and $X \subseteq C_2$. \\
\noindent
Further when ${\cal T}_1$ is compatible with ${\cal T}_2$, we have: \\
\noindent
{\bf 2.} For every (matched) node $l$ in $M_1-X$ of ${\cal T}_1$, its corresponding
matched node $l'$ in ${\cal T}_2$ is present outside the subtree rooted at $n_2$. Analogously, for every (matched) node $r'$ in
$M_2-X$ of ${\cal T}_2$, its corresponding matched node $r$ in ${\cal T}_1$ is present outside the subtree rooted at $n_1$. \\
\noindent
{\bf 3.} If $n_1$ [resp. $n_2$] is a Q-node, then in its child ordering, no node of
$C_1-X$ [resp. $C_2-X$] can be present between two nodes of $X$. \\
\noindent
{\bf 4.}  If $n_1$ and $n_2$ are Q-nodes, then in the child ordering of $n_1$ and
$n_2$, nodes of $X$ appear in the same relative order i.e.~for any three nodes $x_1, x_2, x_3 \in X$, $x_1$ appears between $x_2$
and $x_3$ in the child ordering of $n_1$ if and only if $x_1$ also appears between $x_2$ and $x_3$ in the child ordering of $n_2$. \\
\noindent
{\bf 5.} If $C_1-X$ (resp.$C_2-X$) is non-empty then $U(n_1) \subseteq U(n_2)$ (resp. $U(n_2) \subseteq U(n_1)$).  Further,
 if $C_1-X$ is non-empty then so is $C_2-X$ and hence $U(n_1) = U(n_2)$. \\
\noindent
{\bf 6.} Let $C_1-X$ be non-empty. If $n_1$ [resp.~$n_2$] is a Q-node, then in
its child-ordering either all nodes of $C_1-X$ [resp.~$C_2-X$] appear before the nodes of $X$ or they all appear after the nodes of $X$.
\end{lemma}

\begin{proof}

(1) If there exists an unmatched child $c$ of $n_1$, then as $U(c) \supseteq U(n_1)$, $MMD(c) \subseteq MMD(n_1)$
and $c$ has a greater depth than $n_1$, we would have chosen $c$ over $n_1$. Thus every node in $C_1$ is matched and hence
by the definition of MM-descendants $C_1 = M_1$.

  For the second part, suppose there exists a node $x \in X$ that is not a child of $n_2$. Let $c_2$ be the child of $n_2$ that
contains $x$ as a descendant. $c_2$ must be an unmatched node. (Otherwise $MMD(n_2)$ would have contained $c_2$
and not $x$). But then we would have picked $c_2$ over $n_2$. \\

(2) Let $l$ be a (matched) node in $M_1-X$ (in ${\cal T}_1$) such that the corresponding matched node $l'$ in ${\cal T}_2$ is a
descendant of $n_2$. Note that $l'$ cannot be a child of $n_2$. Otherwise $l' \in M_2$ and thus $l=l'$ is in $X$. Let $p'$ be the parent
of $l'$. Now $p'$ cannot be a matched node. (Otherwise $p'$ would have been matched to $n_1$, a contradiction that $n_1$ is unmatched).
Also $p'$ is a descendant of $n_2$ and hence $U(p') \supseteq U(n_2)$, $MMD(p') \subseteq MMD(n_2)$ and $p'$ has greater depth than
$n_2$. Further $l=l'$ is a common MM-descendant of $n_1$ and $p'$. This contradicts the choice of $n_2$.
   
      Now let $r'$ be a (matched) node in $M_2-X$ (in ${\cal T}_2$), such that the corresponding matched node $r$ is a descendant
of $n_1$. Note that $r$ is not a child of $n_1$, otherwise $r=r'$ is a common MM-descendant of $n_1$ and $n_2$ and hence $r'=r \in X$.
Let $p$ be the parent of $r$ in ${\cal T}_1$. Since $p$ is a
proper descendant of $n_1$, $p$ is a matched node. Let $p$ be matched to a node $p'$ in ${\cal T}_2$. Now $p'$ is a parent of
$r'$ and a descendant of $n_2$. But then the MM-descendants of $n_2$ should not have contained $r'$. \\

(3) Suppose in the child ordering of $n_1$, node $y \in C_1-X$ is present between nodes $x_a \in X$ and $x_b \in X$. Let
$Y, X_a$ and $X_b$ be any max-cliques that are descendants of $y, x_a$ and $x_b$ respectively. Then in any ordering of ${\cal T}_1$,
$Y$ appears between $X_a$ and $X_b$. But by (2), the corresponding matched node $y'$ of $y$ in ${\cal T}_2$ appears outside the
subtree rooted at $n_2$. Thus in any ordering of ${\cal T}_2$, $Y$ appears either before or after both $X_a$ and $X_b$. Thus
${\cal T}_1$ and ${\cal T}_2$ are not compatible. This shows the claim for $n_1$. The proof for $n_2$ is similar. \\

(4) This follows from the fact that ${\cal T}_1$ and ${\cal T}_2$ are compatible and by observing that each matched node (in
particular any node in $X$) contains a max-clique as a descendant. \\

(5) Let $x_a \in X$ be a common child of $n_1$ and $n_2$. Let $X_a$ be a max-clique descendant of $x_a$. Suppose $C_1-X$ is non-empty.
Then let $Y_a$ be any max-clique descendant of a node in $C_1-X$. Note that by (2), $Y_a$ is present outside the subtree rooted at $n_2$.
Now by Lemma \ref{preliminaries}.0 and observing that the least common ancestor of $X_a$ and $Y_a$ (in ${\cal T}_2$) is an ancestor
of $n_2$, we get $U(n_2) \supseteq X_a \cap Y_a \supseteq U(n_1)$. Thus $U(n_1) \subseteq U(n_2)$. Using an analogous argument we can
show that if $C_2-X$ is non-empty then $U(n_2) \subseteq U(n_1)$. This proves the first part of the property.

   For the second part we once again assume that $C_1-X$ is non-empty and hence $U(n_1) \subseteq U(n_2)$. Now if $C_2-X$ is empty
then $MMD(n_2) = X \subset MMD(n_1)$. But this contradicts the choice of $n_1$ (we would have selected $n_2$ instead). \\

(6) By (5), $C_2-X$ is non-empty and $U(n_1) = U(n_2)$. Let $x_a \in X$ and suppose $y_a, y_b \in C_1 - X$ are any two nodes
on different sides of $X$. Let $z_a \in C_2-X$. Note that by (2), the matched nodes of $y_a, y_b$ in ${\cal T}_2$ appear outside the
subtree rooted at $n_2$ and the matched node of $z_a$ in ${\cal T}_1$ appears outside the subtree rooted at $n_1$. Now let
$X_a, Y_a, Y_b$ and $Z_a$ be any descendant max-cliques of $x_a, y_a, y_b$ and $z_a$ respectively. In any leaf-ordering of ${\cal T}_1$,
$X_a$ appears between $Y_a$ and $Y_b$, and $Z_a$ doesn't appear between $Y_a$ and $Y_b$. But in any leaf-ordering of ${\cal T}_2$,
either $Z_a$ and $X_a$ both appear between $Y_a$ and $Y_b$ or they both appear before or after $Y_a$ and $Y_b$. This contradicts
that ${\cal T}_1$ and ${\cal T}_2$ are compatible. Therefore all nodes of $X$ appear before or after all nodes of $C_1 - X$ in the
child ordering of $n_1$. Similarly, the claim also holds for the child ordering of $n_2$ in ${\cal T}_2$.  

\end{proof}

  We now describe the main step of the algorithm. Let $n_1, n_2, M_1, M_2, C_1, C_2$ and $X$ be as defined in the above Lemma. We have four cases
depending on whether $n_1$ and $n_2$ are
P or Q-nodes. In each of these cases, we make progress by either matching two previously unmatched essential nodes of ${\cal T}_1$ and ${\cal T}_2$
or by reducing ${\cal T}_1$ and/or ${\cal T}_2$ at $n_1$ or $n_2$ while preserving their compatibility. We show that our algorithm
requires at most $O(n {\log n})$ iterations and each iteration takes $O(n)$ time. Thus our algorithm runs in $O(n^2 \log{n})$ time.

During the course of the algorithm we may also insert subcliques into a Q-node when we are trying to match it to another Q-node. This is potentially
dangerous as it may destroy the validity of the PQ-tree. When the Q-nodes have the same universal set, this trouble does not arise. However, in
case the two Q-nodes have different universal sets, we need to re-structure the trees. Case 4, when $n_1$ and $n_2$ are both Q-nodes, has subcases to
deal with these complications.   \\

\begin{figure}
\begin{center}
\subfigure[$C_1-X$ is empty] {
    \label{pp1}
    \includegraphics[scale=.38]{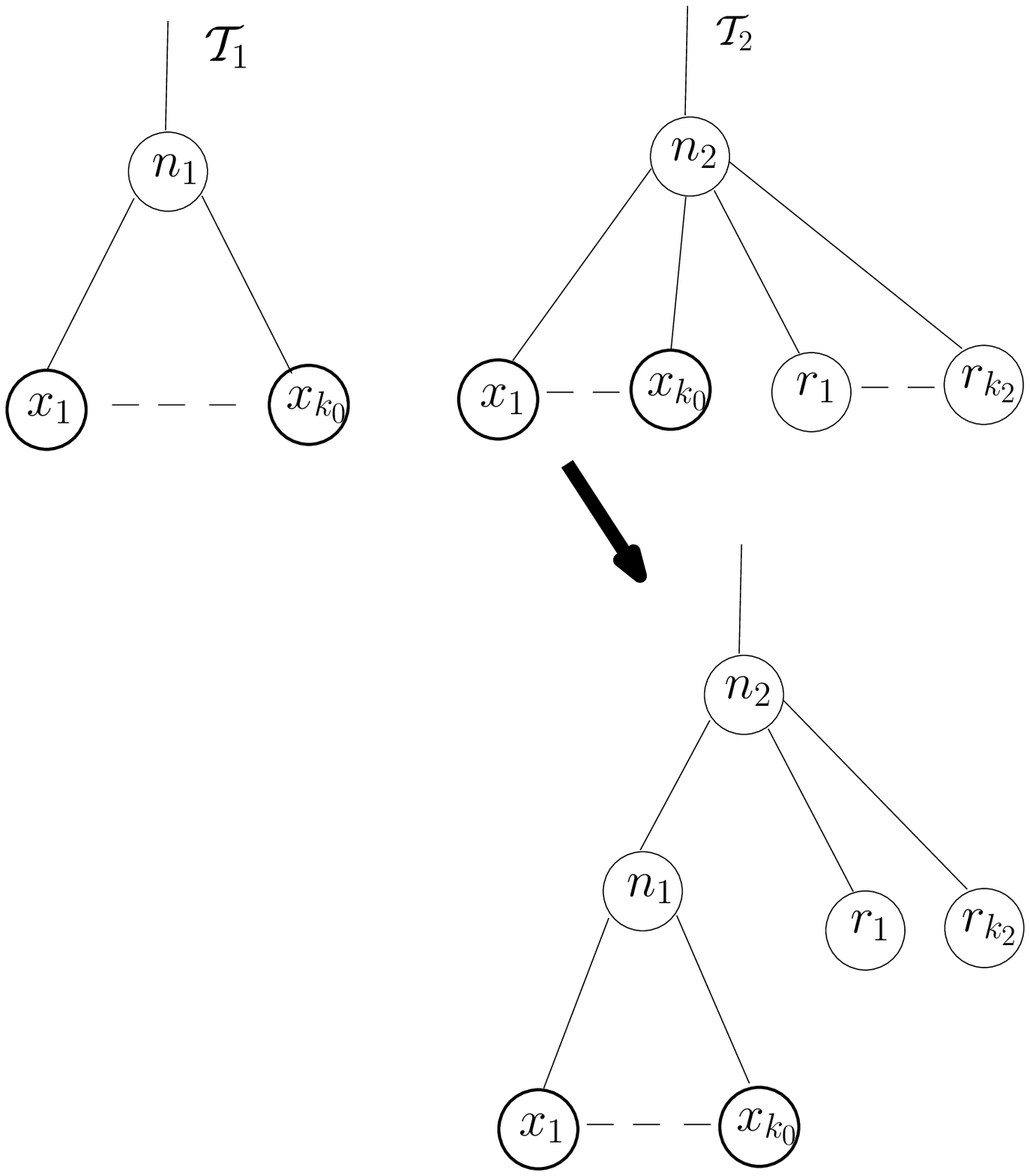}}\hspace{1cm}
\subfigure[$C_1-X$ is non-empty] {
    \label{pp2}
    \includegraphics[scale=.38]{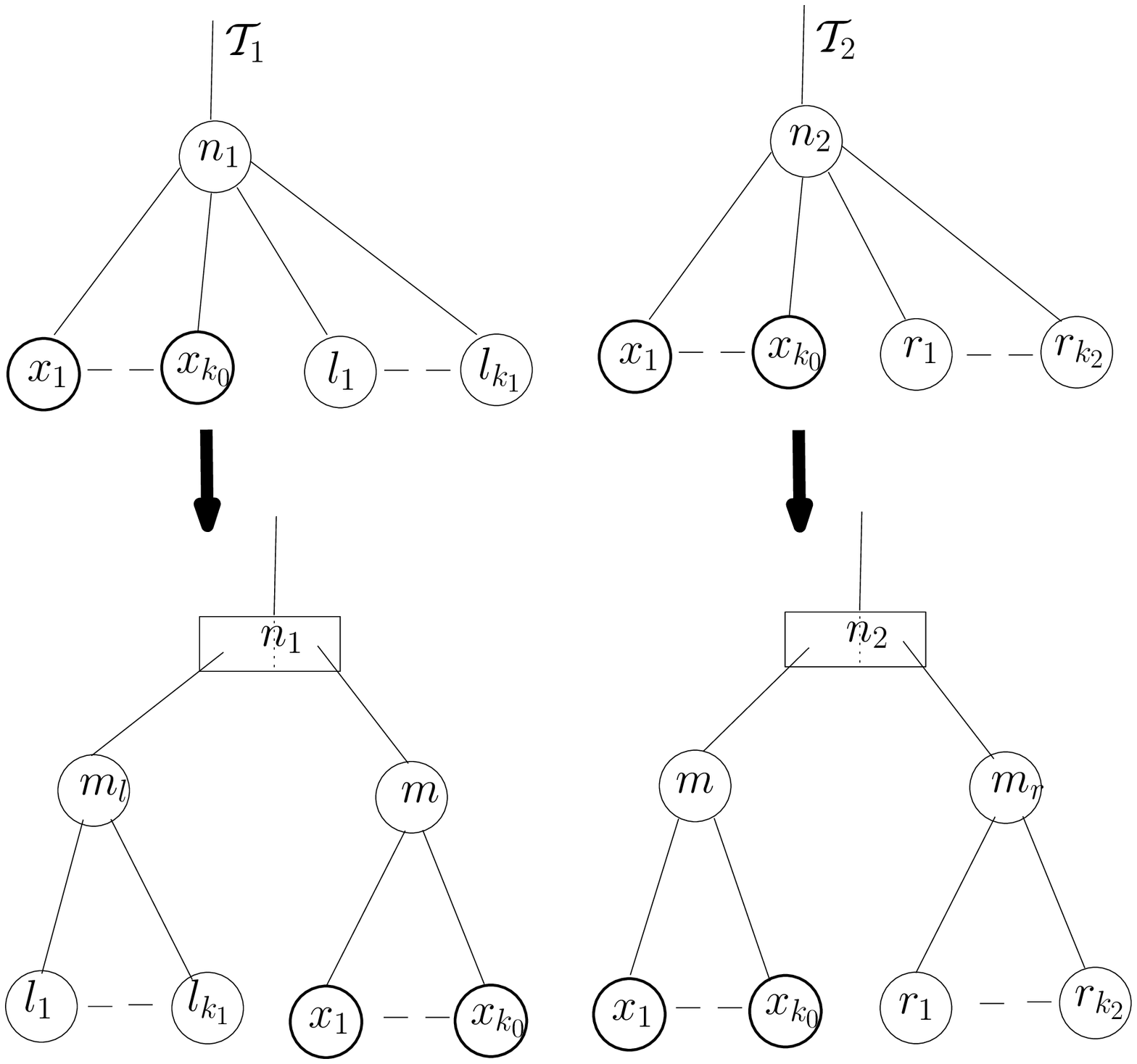}}\hspace{1cm}
\caption{Reduction templates for Case 1}
\end{center}
\end{figure}

\noindent
\textbf{\large Case 1: $n_1$ and $n_2$ are both P-nodes.} \\
  By Corollary \ref{simplification}, the children of $n_1$ and $n_2$ are essential nodes, so $C_1$ and $C_2$ are precisely the children of
$n_1$ and $n_2$ respectively. Let $X$ consist of nodes $\brac{x_1, \cdots, x_{k_0}}$. If $C_2-X$ is empty, then by Lemma
\ref{properties}.5, $C_1-X$ is also empty and hence $n_1$ and $n_2$ are the same. So we match $n_1$ with $n_2$ and go to the next iteration.
Suppose now that $C_2-X$ is non empty. Let $C_2 - X = \brac{r_1, \cdots, r_{k_2}}$. If $C_1-X$ is empty, then we use the reduction
template of Figure \ref{pp1} to modify ${\cal T}_2$, matching the new parent of $X$ in ${\cal T}_2$ to $n_1$. It is easy to see that
${\cal T}_1$ is compatible with ${\cal T}_2$ if and only if ${\cal T}_1$ is compatible with the modified ${\cal T}_2$.

    Now let $C_1-X = \brac{l_1, \cdots, l_{k_1}}$ be non-empty. In this case we use the
reduction template of Figure \ref{pp2} to modify ${\cal T}_1$ and ${\cal T}_2$ to ${\cal T}'_1$ and ${\cal T}'_2$ respectively.
Note that it is possible to have $k_i=1$ for some $i$'s, in which case the template is slightly different because we do not make a
node with one child, however, the reduction always makes progress as each $n_i$ has at least 3 children.

We now claim that ${\cal T}_1$ is compatible with ${\cal T}_2$ if and only if ${\cal T}'_1$ is compatible with ${\cal T}'_2$. The
reverse direction is trivial. For the forward direction, let ${\cal O}_1$ and ${\cal O}_2$ be two compatible leaf orderings of
${\cal T}_1$ and ${\cal T}_2$ respectively. Recall that by Lemma \ref{properties}.2, for every [matched] node of $C_1-X$ in ${\cal T}_1$,
the corresponding matched node in ${\cal T}_2$ appears outside the subtree rooted at $n_2$. This implies that the descendant nodes
of $\brac{x_1, x_2, \cdots, x_{k_0}}$ all appear consecutively in ${\cal O}_1$. Hence the descendant nodes of
$\brac{x_1, x_2, \cdots, x_{k_0}}$ also appear consecutively in ${\cal O}_2$. Thus we conclude that ${\cal T}_1$
and ${\cal T}_2$ are compatible if and only if the reduced trees ${\cal T}'_1$ and ${\cal T}'_2$ are also compatible.
Note that both the template reductions take at most $O(n)$ time. \\

\noindent
\textbf{\large Case 2: $n_1$ is a P-node and $n_2$  is a Q-node.} \\
    If $C_1-X = \emptyset$, we reduce ${\cal T}_1$ by ordering the children of $n_1$ as they appear in the child ordering of $n_2$, and changing
$n_1$ into a Q-node (and leading to Case 4). This reduction preserves the compatibility of the two trees.

\begin{figure}
\begin{center}
\scalebox{0.55}{\includegraphics{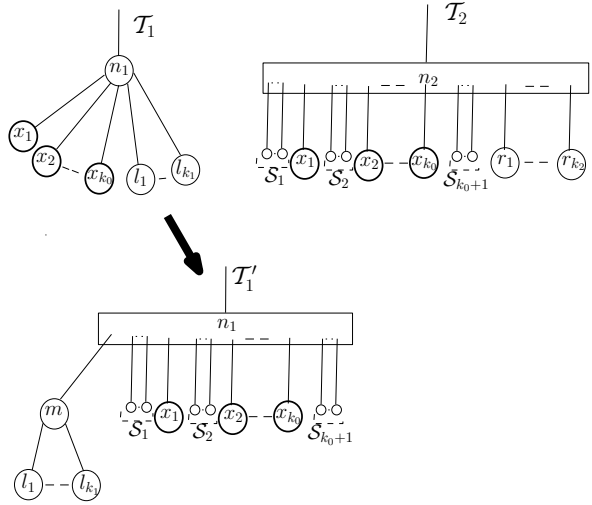}}
\end{center}
\caption{Reduction template for Case 2, when $C_1-X \not= \emptyset$}
\label{pq1}
\end{figure}

    Now suppose $C_1-X \not= \emptyset$. Lemma \ref{properties}.5 implies that, $C_2-X \not= \emptyset$ and $U(n_1) = U(n_2)$.
By Lemma \ref{properties}.6, we can assume that the nodes in $X$ appear before the nodes in $C_2-X$ in the child ordering of $n_2$.
Now let $X = {x_1, \cdots, x_{k_0}}$, $C_1-X = {l_1, \cdots,l_{k_1}}$ and $C_2-X = {r_1, \cdots, r_{k_2}}$.
For $i \in {2,\cdots,k_0}$,
let ${\cal S}_i$ be the sequence of subcliques that appear between $x_{i-1}$ and $x_i$ in the child ordering of $n_2$. Note that
${\cal S}_i$ consists of the right tail of $x_{i-1}$ followed by the left tail of $x_{i}$. We let ${\cal S}_1$ and ${\cal S}_{k_0+1}$
denote the left and right tails of $x_1$ and $x_{k_0}$ respectively. We now reduce the subtree rooted at $n_1$ as shown in Figure
\ref{pq1}, changing it into a Q-node. Clearly $U(n_1)$ is preserved in this operation. The correctness of 
this operation follows by Lemma \ref{properties}.2. It is easy to see that both the template reductions run in $O(n)$ time. \\

\noindent
\textbf{\large Case 3: $n_1$ is a Q-node and $n_2$  is a P-node.} \\
  If $C_2-X$ is empty, then we reduce ${\cal T}_2$ by ordering the child nodes of $n_2$ (i.e.~$X$) as they appear in the child ordering of $n_1$,
and changing $n_2$ into a Q-node.

  Now let $C_2-X$ be nonempty. By Lemma \ref{properties}.5, $U(n_2) \subseteq U(n_1)$. Let $X = \brac{x_1, x_2, \cdots, x_{k_0}}$,
$C_2-X = \brac{r_1, \cdots, r_{k_2}}$ and ${\cal S}_1, \cdots,$
${\cal S}_{k_0+1}$ be defined as in the previous case: ${\cal S}_1$ is the left tail of $x_1$ (in ${\cal T}_1$), ${\cal S}_i$ is the
concatenation of the right tail of $x_{i-1}$ and the left tail of $x_i$, for $i \in \brac{2,\cdots,k_0}$ and ${\cal S}_{k_0+1}$ is the
right tail of $x_{k_0}$.

\begin{figure}
\begin{center}
\scalebox{0.6}{\includegraphics{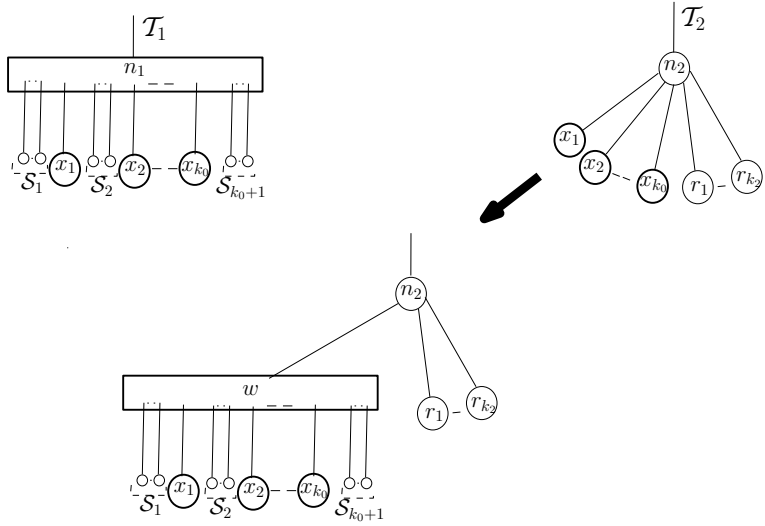}}
\end{center}
\caption{Reduction template for Case 3, when $n_1$ is a Q-node, $n_2$ is a P-node and $C_1-X$ is empty.}
\label{qp1}
\end{figure}

  Now if $C_1-X$ is empty, then we use the template of Figure \ref{qp1} to reduce ${\cal T}_2$, grouping all nodes of $X$ into a new Q-node $w$,
ordering them in the way they appear in ${\cal T}_1$ and inserting the subclique children of $n_1$ into $w$. Note that since
$U(n_2) \subseteq U(n_1)$, this operation doesn't change $U(n_2)$ and hence it preserves the validity of ${\cal T}_2$. Further $n_1$
is identical to $w$ and hence we match these nodes. Thus we make progress even when $|X| = 1$.

  If $C_1-X$ is non-empty, we use the template similar to Figure \ref{pq1} (to reduce ${\cal T}_2$) in which the roles of $n_1$ and $n_2$ have
been switched. Note that the template reductions of this case run in $O(n)$ time.\\

\noindent
\textbf{\large Case 4: $n_1$ and $n_2$ are both Q-nodes} \\
  Let $X = \brac{x_1, \cdots, x_{k_0}}$ appear in that order in the child ordering of $n_1$ and $n_2$. (They appear in the same order because of
Lemma \ref{properties}.4.) Let $p_1$ and $p_2$ be the parents of $n_1$ and $n_2$ respectively.

 If $n_1$ and $n_2$ have no other children than $X$, we match $n_1$ with $n_2$ and proceed to the
next iteration. More typically, they have other children. These may be essential nodes to one side or the other of $X$ (by Lemma \ref{properties}.6)
or subclique nodes interspersed in $X$ as tails of the nodes of $X$. We give a high-level outline of Case 4, beginning with a discussion of
subclique nodes.

 For $i \in \brac{1,\cdots, k_0}$,
let ${\cal L}_i$ and ${\cal R}_i$ be the left and right tails of $x_i$ in ${\cal T}_1$ and, ${\cal L}'_i$ and ${\cal R}'_i$ be the
left and right tails of $x_i$ in ${\cal T}_2$. The only way to deal with the subclique nodes is to do subclique insertions in both trees to
merge the tails. This is because in any intersection tree ${\cal T}_I$ obtained from ${\cal T}_1$ and ${\cal T}_2$, the tails of $x_i$ in ${\cal T}_I$ must contain the merge of the tails of $x_i$ in ${\cal T}_1$ and ${\cal T}_2$. So long as $|X| \ge 2$, the ordering $x_1, \cdots, x_{k_0}$ completely
determines which pairs of tails must merge: ${\cal L}_i$ must merge with ${\cal L}'_i$ and ${\cal R}_i$ must merge with ${\cal R}'_i$.

The case $|X|=1$ is more complicated because the intersection tree may merge ${\cal L}_1$ with ${\cal L}'_1$ and ${\cal R}_1$ with ${\cal R}'_1$
or merge ${\cal L}_1$ with $\bar{\cal R}_1$ and ${\cal R}_1$ with $\bar{\cal L}_1$. This decision problem is referred to as the {\em alignment problem}. We prove (at the beginning of Case 4.3) that in case both choices give mergable pairs, then either choice yields an intersection tree,
if an intersection tree exists.

This completes our high-level discussion of subclique nodes. We continue with a high-level description of the subcase structure for Case 4.
We have subcases depending on whether $U(n_1) = U(n_2)$ and whether $n_1$ and $n_2$ have the same essential children.
If both these conditions hold, then we merge the tails of the nodes of $X$ and match $n_1$ with $n_2$. (In other words we replace ${\cal L}_i$ and 
${\cal L}'_i$ with ${\cal L}_i + {\cal L}'_i$, and replace ${\cal R}_i$ and ${\cal R}'_i$ with ${\cal R}_i + {\cal R}'_i$). The cost of matching 
any two nodes $x$ and $y$ is  $(m_x + m_y) |I|)$, where $m_x$ and $m_y$ are the number of subclique children of $x$ and $y$ respectively. Once a 
node is matched its subclique children will not change. Hence the total amortized cost of matching all the nodes is $O(n \cdot |I|) = O(n^2)$.

When $U(n_1) \not= U(n_2)$ or when $n_1$ and $n_2$ do not have the same essential children then we have three subcases.
Case 4.1 handles the situation when $U(n_1) \not\supseteq U(n_2)$. In this case we either insert subcliques
of one tree into another and match $n_1$ with $n_2$ or we do some subclique insertions that will take us to the case when $U(n_1) \supseteq U(n_2)$. 
The remaining cases handle the situation when $U(n_1) \supseteq U(n_2)$, Case 4.2 when $C_1-X$ is non-empty and Case 4.3 when it is empty.
In both cases, we reduce ${\cal T}_1$ but the details vary. However in both cases our reduction templates depend on
whether  $p_1$ is a P-node or a Q-node. If $p_1$ is a P-node, we reduce ${\cal T}_1$ by grouping some of the child nodes of $p_1$
into a single node, deleting them and adding the node as a first or last child of $n_1$. If $p_1$ is a Q-node then there are two ways of reducing:
delete $n_1$ and reassign its children as children of $p_1$ or reverse the children of $n_1$, delete $n_1$ and reassign its children as children of
$p_1$. We refer to this operation as a {\em collapse}. We now give the details of each case. \\

\noindent
\textbf{\large Case 4.1: $U(n_1) \not\supseteq U(n_2)$} \\
   Since $n_1$ was chosen so that $U(n_1)$ is maximal, we also have $U(n_2) \not\supseteq U(n_1)$. Now using Lemma \ref{properties}.5, we infer
that $C_1-X$ is empty and $C_2-X$ is empty. Thus the difference between $U(n_1)$ and $U(n_2)$ arises due to the subcliques. Let $L$ be a subclique
that is either the first or the last child of $n_1$, with the property that $L \not\subseteq U(n_2)$. Such a subclique
exists since by Lemma \ref{preliminaries}.6, the intersection of the universal sets of the first and last child nodes of $n_1$ is $U(n_1)$.
Also let $R$ be a subclique that is either the first or the last child of $n_2$, with the property that $R \not\subseteq U(n_1)$.

   Note that even if $|X|=1$, the alignment is unique since $L$ and $R$ cannot appear in the same tail of $x_1$ in any intersection tree.
Further, we can assume without loss of generality that $L$ is present in the left tail of $x_1$ in ${\cal T}_1$ and $R$ is present in the
right tail of $x_{k_0}$ in ${\cal T}_2$.

   Let $X_1$ be any max-clique descendant of $x_1$. If $p_1$ is a P-node then we claim that $U(p_1) \subseteq U(n_2)$. To see this, let $Z$
be a max-clique descendant of $p_1$, that is not a descendant of $n_1$.  In ${\cal T}_2$, $Z$ appears outside the subtree rooted at $n_2$.
Now by applying Lemma \ref{preliminaries}.0
on $X_1$ and $Z$, we conclude that every descendant of $n_2$ must contain the vertex set $Z \cap X_1$. Thus we have
$U(p_1) \subseteq (Z \cap X_1) \subseteq U(n_2)$. Note that if ${\cal T}_1$ and ${\cal T}_2$ are compatible, then in any intersection
tree of ${\cal T}_1$ and ${\cal T}_2$, the nodes of ${\cal L}_1$ and ${\cal L}'_1$ appear in the left tail of $x_1$ and the nodes of
${\cal R}_{k_0}$ and ${\cal R}'_{k_0}$
appear in the right tail of $x_{k_0}$.  Now if ${\cal L}'_1$ is non-empty, then we insert the left most subclique of ${\cal L}'_1$
into ${\cal L}_1$ (at the appropriate location so that the resulting sequence is still a subclique ordering), as a child of $n_1$. Also if
${\cal R}'_{k_0}$ is non-empty, then we insert the right most subclique of ${\cal R}'_{k_0}$ into ${\cal R}_{k_0}$, as a child of $n_1$.
These insertions change $U(n_1)$ to $U(n_1) \cap U(n_2) \supseteq U(p_1)$ and we would be in case 4.3 with the roles of
$n_1$ and $n_2$ being reversed. (Note that since the universal set of the modified $n_1$ is  a superset of the universal set of $p_1$, the resulting reduced tree of ${\cal T}_1$ is valid). Although this doesn't constitute a progress step since the number of leaf orderings of $n_1$ doesn't
change, we will make progress in the Case 4.3.

   Similarly if $p_2$ is a P-node then we insert the first subclique of ${\cal L}_1$ (if it exists) into ${\cal L}'_1$ and the last subclique of
${\cal R}_{k_0}$ (if it exists)  into ${\cal R}'_{k_0}$. After this we would be in Case 4.3.
 
   Now if the parents of $n_1$ and $n_2$ are both Q-nodes then we look at the tails of $n_1$ and $n_2$. If all the subcliques in these
tails are subsets of $U(n_1) \cap U(n_2)$, then we replace ${\cal L}_i$ and ${\cal L}'_i$ with ${\cal L}_i + {\cal L}'_i$ and ${\cal R}_i$
and ${\cal R}'_i$ with ${\cal R}_i + {\cal R}'_i$. This changes $U(n_1)$ and $U(n_2)$ to $U(n_1) \cap U(n_2)$ and makes $n_1$ identical
to $n_2$. Thus we match $n_1$ with $n_2$ and iterate.

   Otherwise without loss of generality let the subclique $S \not\subseteq U(n_2)$ be present in the (say left) tail of $n_1$. Observe that in any
intersection tree $S$ and $R$ cannot be present in the same tail of $x_{k_0}$ (since neither is a subset of the other). This implies
that we can reduce the tree ${\cal T}_1$ by collapsing $n_1$ i.e.~by removing $n_1$, inserting the sequence of child nodes of $n_1$ after $S$
($S$ and $L$ are now in the left tail of $x_1$), and assigning $p_1$  as their parent. This completes case 4.1. Note that all
the steps in this case take $O(n)$ time, except the matching step (recall that all the matching steps take $O(n^2)$ amortized time). \\

\noindent
\textbf{\large Case 4.2: $U(n_1) \supseteq U(n_2)$ and $C_1-X$ is non-empty.} \\
 By Lemma \ref{properties}.5, $C_2-X$ is also non-empty and further $U(n_1)$ is equal to $U(n_2)$. In this case we will reduce ${\cal T}_1$
depending on whether $p_1$ is a P-node or a Q-node. Further when $p_1$ is a Q-node, our reduction template also depends on whether $n_1$
has sibling essential nodes.

Let $l_1, l_2, \cdots, l_{k_1}$ be the essential nodes in $C_1-X$ appearing in that order and appearing (without loss of generality)
before the nodes of $X$ in ${\cal T}_1$. Note that by Lemma \ref{properties}.2, for each node in $C_1-X$, the corresponding matched node in
${\cal T}_2$ appears outside the subtree rooted at $n_2$. Thus if ${\cal T}_1$ and ${\cal T}_2$ are compatible, then all the nodes of
$C_2-X$ must appear after the nodes of $X$ in the child ordering of $n_2$. Let these nodes be $r_1, r_2, \cdots, r_{k_2}$. \\

\noindent
\textbf{\large Case 4.2.1: $p_1$ is a P-node} \\
 Let $Y = \brac{y_1, y_2, \cdots, y_{k_3}}$ be the child nodes of $p_1$ other than $n_1$ . Also, let ${\cal T}_I$ be any intersection tree of
${\cal T}_1$ and ${\cal T}_2$.  We first observe that for $i \in \brac{1, \cdots, k_0}$ and $j \in \brac{1, \cdots, k_2}$, 
$MMD(y_i) \cap MMD(r_j) \not= \emptyset$, if and only if $y_i$ and $r_j$ have a max-clique descendant.

\begin{figure}
\begin{center}
\scalebox{0.7}{\includegraphics{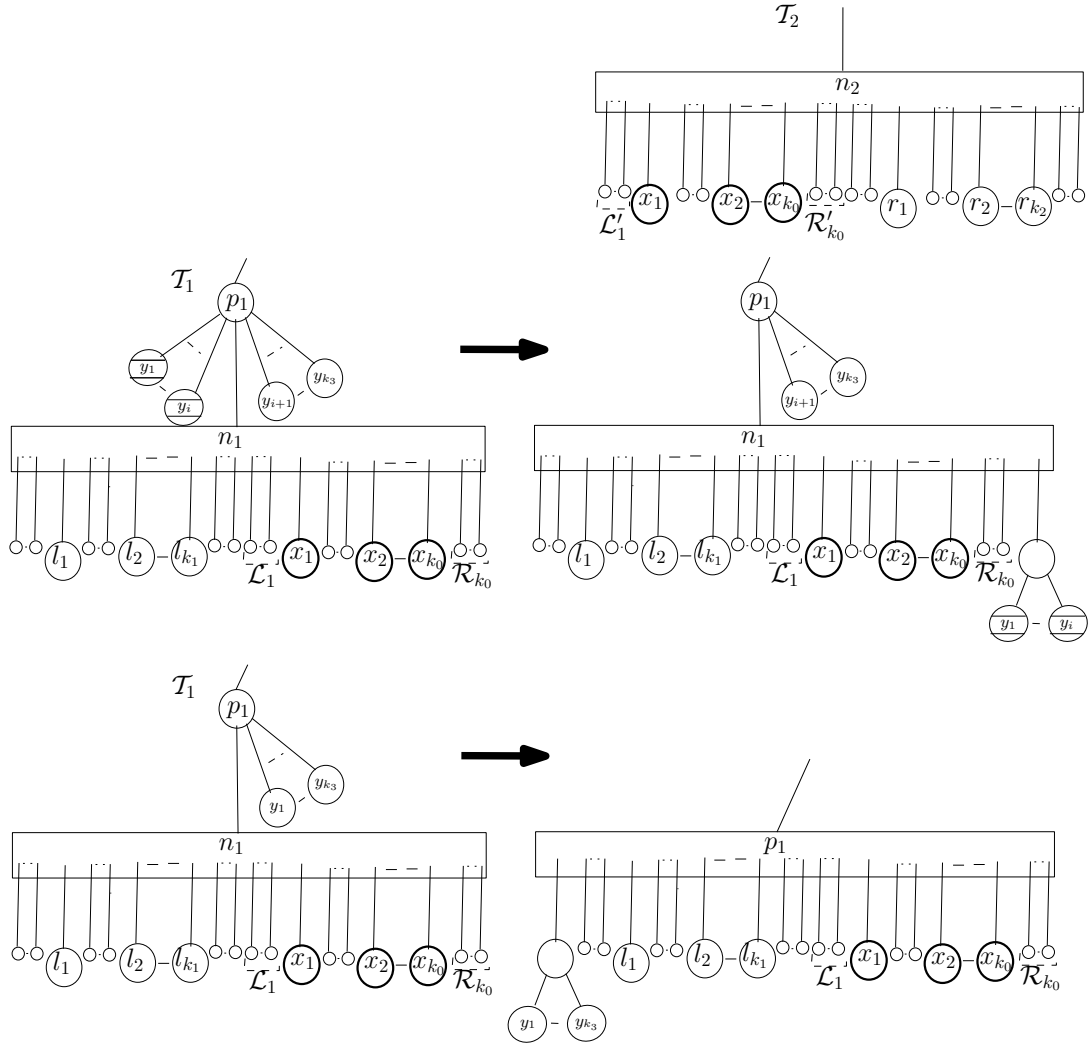}}
\end{center}
\caption{Reduction template of ${\cal T}_1$ for Case 4.2.1. A node $y_a$ has horizontal stripes if
$MMD(y_a) \cap MMD(r_b) \not= \emptyset$ for some $r_b$ and no stripes otherwise.}
\label{qq6}
\end{figure}

For any such pair $y_i$ and $r_j$, let $MMD(y_i) \cap MMD(r_j) \not= \emptyset$ and let $Y$ be a common max-clique descendant of $y_i$ and 
$r_j$. Then note that because of the constraints imposed by
the child ordering of $n_2$, in any leaf ordering of ${\cal T}_I$, the descendant cliques of $l_1$ do not appear between the descendant
cliques of $x_1$ and $Y$. Thus $y_i$ must appear after $x_{k_1}$, and so we reduce ${\cal T}_1$, by grouping all nodes $y_i$ satisfying
$MMD(y_i) \cap MMD(r_j) \not= \emptyset$ for some $r_j$ into a P-node and adding it as a child node of $n_1$ to the (immediate) right of
${\cal R}_{k_0}$ as shown in Figure \ref{qq6}(top).

  Now if $MMD(y_i) \cap MMD(r_j) = \emptyset$ for all $y_i$ and $r_j$, then the above reduction doesn't apply. But in this case (because of
the constraints on $n_2$), for every $y_i$ and every leaf ordering of ${\cal T}_I$, no max-clique descendant of $y_i$ appears between
the max-clique descendants of $n_2$. Thus we group all the nodes of $Y$ into a P-node and add it as a child of $n_1$ to the left of $l_1$
as shown in Figure \ref{qq6}(bottom).

Note that for any two distinct nodes $y_a, y_b \in \brac{y_1, \cdots, y_{k_3}}$ we have:
$MMD(y_a) \cap MMD(y_b) = \emptyset$. Similarly, for any two distinct nodes $r_a, r_b \in \brac{r_1, \cdots, r_{k_2}}$ we have:
$MMD(r_a) \cap MMD(r_b) = \emptyset$. This implies that we can first compute the MM-Descendants of all $y_i$ and $r_j$ in $O(n)$ 
time and further we can compute all nodes $y_i$ that satisfy $MMD(y_i) \cap MMD(r_j) \not= \emptyset$ for some $r_j$, in $O(n)$ 
time. Thus the template reductions of Figure \ref{qq6} run in $O(n)$ time. \\

\noindent
\textbf{\large Case 4.2.2: $p_1$ is a Q-node and $n_1$ is its only essential child.} \\
 Since the only essential child of $p_1$ is $n_1$, all of its remaining children are subcliques that are present as tails of $n_1$. Thus each
of these subcliques is a subset of $U(n_1)$.
 Now let $Z$ and $R$ be any two max-clique descendants of $x_{k_0}$ and $r_1$ respectively.
By Lemma \ref{properties}.2, $R$ appears outside the subtree rooted at $n_1$ (in ${\cal T}_1$) and hence outside the subtree rooted at $p_1$.
By Lemma \ref{preliminaries}.0, we conclude that each descendant clique of $p_1$ must contain $Z \cap R$. Thus we have
$U(p_1) \supseteq Z \cap R \supseteq U(n_2)=U(n_1) \supseteq U(p_1)$. Hence all of these sets must be equal and hence we infer the following:
$Z \cap R = U(n_1)$ and hence $U(x_{k_0}) \cap U(r_1) = U(n_1)$. Further, each subclique child of $p_1$ must precisely be the clique $U(n_1)$.

\begin{figure}
\begin{center}
\scalebox{0.6}{\includegraphics{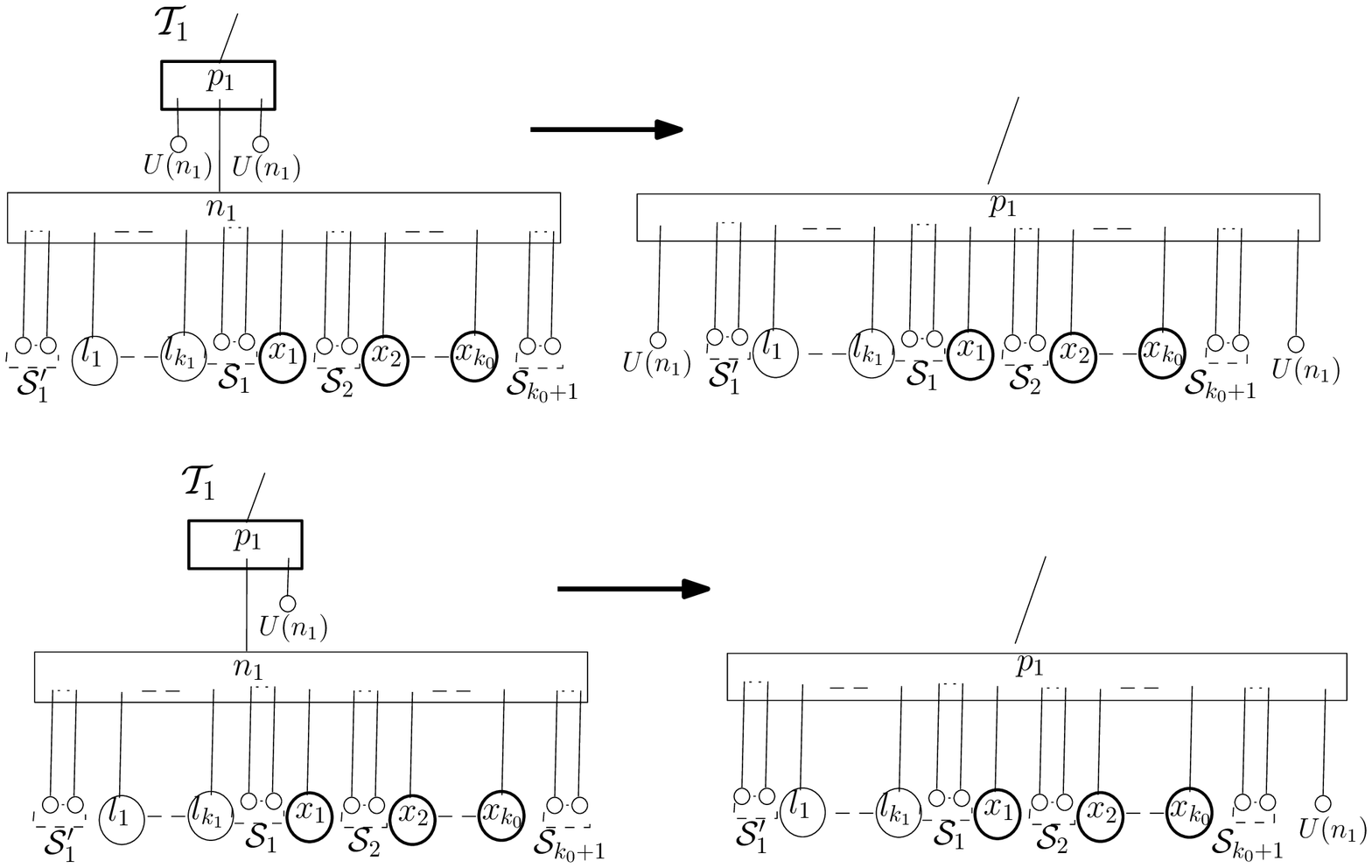}}
\end{center}
\caption{Reduction templates of ${\cal T}_1$ for Case 4.2.2.}
\label{qq1}
\end{figure}

Since we have eliminated adjacent duplicates from all Q-nodes, there can be at most one such subclique
in each tail of $n_1$. Now if the subclique ($U(n_1)$) appears on both sides of $n_1$, then there is a unique way of collapsing $n_1$
(see Figure \ref{qq1}(top)). Otherwise we collapse $n_1$ in such a way that $U(n_1)$ is present in the tail of $x_{k_0}$ as shown in
Figure \ref{qq1}(bottom). This is justified (i.e.~it preserves compatibility between ${\cal T}_1$ and ${\cal T}_2$) because $U(n_1)$ can
be inserted into the right tail of $x_{k_0}$ in both ${\cal T}_1$ and ${\cal T}_2$. In other words, if ${\cal T}_1$ and ${\cal T}_2$ are
compatible, then there exists an intersection tree in which $U(n_1)$ is present in the right tail of $x_{k_0}$. The template reductions
of this case, clearly run in $O(n)$ time.\\

\begin{figure}
\begin{center}
\scalebox{0.6}{\includegraphics{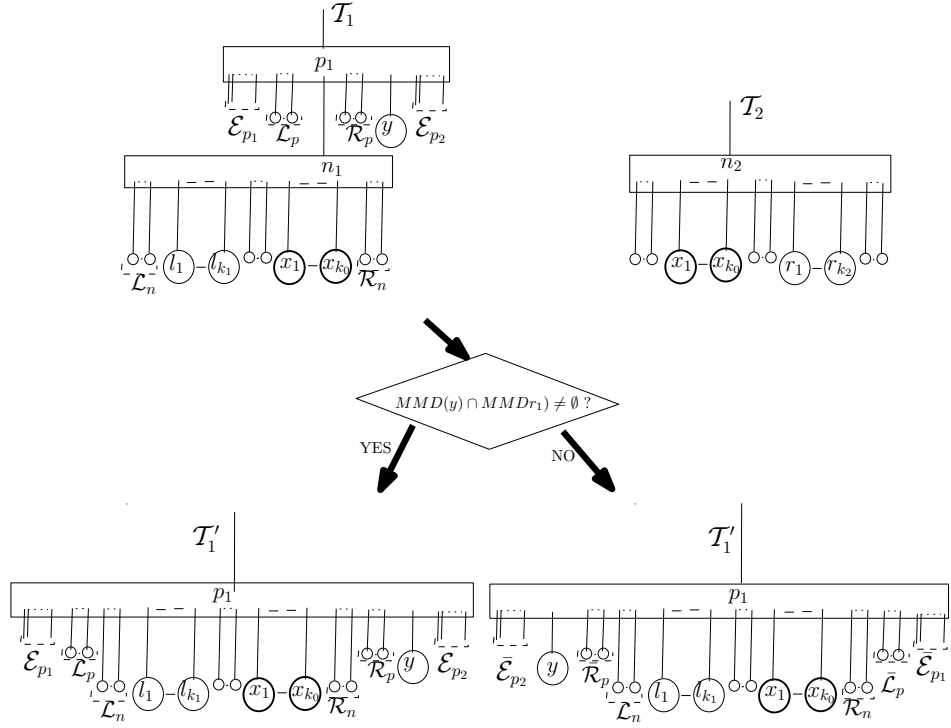}}
\end{center}
\caption{Reduction template of ${\cal T}_1$ for Case 4.2.3}
\label{qq7}
\end{figure}

\noindent
\textbf{\large Case 4.2.3: $p_1$ is a Q-node and has more than one essential child.} \\
 Let $y$ be an essential child of $p_1$, such that all the nodes between $n_1$ and $y$ are subcliques. Without loss of generality, we assume
that $y$ appears to the right of $n_1$. We collapse $n_1$, depending on whether $MMD(y) \cap MMD(r_1)$ is empty or not, as shown in 
Figure \ref{qq7}. Thus the template reduction runs in $O(n)$ time.

If $MMD(y) \cap MMD(r_1)$ is non-empty, there exists a max-clique $Y$ that is a descendant of both
$r_1$ and $y$. Now if ${\cal T}_1$ and ${\cal T}_2$ are compatible, then in the leaf ordering of any intersection tree, the max-clique
descendants of $x_{k_0}$ appear in between the max-clique descendants of $l_1$ and $Y$. Thus we collapse the node $n_1$, by deleting
$n_1$, and reassigning $p_1$ as the parent of all the children of $n_1$. (Thus no essential node appears between $x_{k_0}$ and $y$).

On the other hand if $MMD(y) \cap MMD(r_1)$ is empty, we observe the following: In the leaf ordering of any intersection tree 
${\cal T}_I$ no max-clique appears in between the  max-clique descendants of $x_{k_0}$ and the max-clique descendants of $r_1$. 
Therefore, in this case we collapse $n_1$, by reversing its children, deleting it, and reassigning $p_1$ as the parent of all the 
children of $n_1$. (Thus no essential node appears between $l_1$ and $y$).\\

\noindent
\textbf{\large Case 4.3: $U(n_1) \supseteq U(n_2)$ and $C_1-X$ is empty} \\
   As before we have three cases depending on whether $p_1$ is a P-node or a Q-node and whether $p_1$ has more than
one essential child. In each of these cases, when $|X|=1$, we need to first solve the alignment problem (as a preprocessing step). Also
when $p_1$ is a Q-node, unlike in Case 4.2, both ways of collapsing $n_1$ may lead to a valid intersection tree. \\

\noindent
\textbf{Alignment Problem}

   Recall that when $|X|=1$ (and $C_1-X = \emptyset$), the alignment may not be unique i.e.~one of the following might happen in the
intersection tree ${\cal T}_I$.
\begin{enumerate}
\item Left and right tails of $x_1$ (in ${\cal T}_I$) contain ${\cal L}_1 + {\cal L}'_1$ and ${\cal R}_1 + {\cal R}'_1$ respectively.
\item Left and right tails of $x_1$ (in ${\cal T}_I$) contain $\bar{\cal R}_1 + {\cal L}'_1$ and $\bar{\cal L}_1 + {\cal R}'_1$ respectively.
\end{enumerate}

  If one of the merges in (1) or (2) is invalid, then there is only a single way of aligning the tails, otherwise we show in the following Lemma
that if ${\cal T}_1$ and ${\cal T}_2$ are compatible, then choosing either one of the two alignments will work.

\begin{lemma}
\label{alignment}
Let $U(n_1) \supseteq U(n_2)$, $|X|=1$ and $C_1-X$ be empty. Let ${\cal L}_1$, ${\cal R}_1$ be the left and right tails of $x_1$ in
${\cal T}_1$ and ${\cal L}'_1$, ${\cal R}'_1$ be the left and right tails of $x_1$ in ${\cal T}_2$.  If both ways of alignment are mergable
i.e.~(a) ${\cal L}_1 + {\cal L}'_1$, ${\cal R}_1 + {\cal R}'_1$ are valid and (b) $\bar{\cal R}_1 + {\cal L}'_1$, $\bar{\cal L}_1 + {\cal R}'_1$
are valid, then there exists an intersection tree ${\cal T}_I$ (of ${\cal T}_1$ and ${\cal T}_2$) with ${\cal L}_1 + {\cal L}'_1$ and
${\cal R}_1 + {\cal R}'_1$ contained in the left and right tails of $x_1$ (respectively) if and only if there exists an intersection tree
${\cal T}'_I$ with $\bar{\cal R}_1 + {\cal L}'_1$ and $\bar{\cal L}_1 + {\cal R}'_1$ contained in the left and right tails of $x_1$
(respectively).
\end{lemma}

\begin{proof}
   Let ${\cal L}$, ${\cal R}$ be the left and right tails of $x_1$ in an intersection tree ${\cal T}_I$. Each subclique $S$ in
${\cal L}$ or ${\cal R}$ appears as a subclique in ${\cal T}_1$ or ${\cal T}_2$. In particular we observe the following: \\

\noindent
\textit{Property 1:} If $S$ is a subclique in ${\cal L}$ or ${\cal R}$ then in ${\cal T}_1$ or ${\cal T}_2$, $S$ is present in a tail of $x_1$
or in a tail of an ancestor of $x_1$. \\

  Note that since $n$ contains at least two children, ${\cal L}$ and ${\cal R}$ both cannot be empty. If one of them, say ${\cal L}$ is empty
then the last clique in ${\cal R}$ must be $U(n_1)$. If both ${\cal L}$ and ${\cal R}$ are non-empty then
the intersection of the first subclique of ${\cal L}_1$ with the last subclique of ${\cal R}_1$ is $U(n_1)$.
In either case we observe that, since ${\cal L}_1 + {\cal L}'_1$ and $\bar{\cal R}_1 + {\cal L}'_1$ are both valid (subclique orderings),
each subclique in ${\cal L}'_1$ is either a superset of $U(n_1)$ or a subset of $U(n_1)$. Similarly, since ${\cal R}_1 + {\cal R}'_1$ and
$\bar{\cal L}_1 + {\cal R}'_1$ are both valid (superclique orderings), each subclique in ${\cal R}'_1$ is either a superset of $U(n_1)$ or a subset
of $U(n_1)$. Further for any ancestor $n_a$ of $n_2$, $U(n_a) \subseteq U(n_2) \subseteq U(n_1)$ and hence the tails of any such $n_a$ would
consist of subcliques that are subsets of $U(n_1)$. Note that this condition also holds for any ancestor of $n_1$ in ${\cal T}_1$.

  By above conditions and (1) we infer that for any subclique $S$ in ${\cal L}$ or ${\cal R}$, $S$ is either a superset of $U(n_1)$ or
a subset of $U(n_1)$. Furthermore, if $S$ is a superset of $U(n_1)$ then it is present in one of ${\cal L}_1, {\cal R}_1, {\cal L}'_1$
 or ${\cal R}'_1$.
This implies that if there exists an intersection tree ${\cal T}_I$ in which ${\cal L}$ contains ${\cal L}_1 + {\cal L}'_1$ and $R$
contains ${\cal R}_1 + {\cal R}'_1$, then replacing ${\cal L}$ with ${\cal L}-{\cal L}_1 + \bar{\cal R}_1$ and ${\cal R}$ with
${\cal R}-{\cal R}_1+\bar{\cal L}_1$ also results in a valid intersection tree.
\end{proof}

 Note that the amortized cost of doing the mergability checks (a) and (b) of Lemma \ref{alignment} (over all iterations of the algorithm)
is $O(n \cdot |I|) = O(n^2)$. For the rest of the cases, we can assume that ${\cal L}_1$ is aligned with ${\cal L}_2$ and ${\cal R}_1$ is aligned
with ${\cal R}_2$. In  other words if ${\cal T}_1$ and ${\cal T}_2$ are compatible, then there exists an intersection tree that contains
${\cal L}_1 + {\cal L}_2$ and ${\cal R}_1 + {\cal R}_2$ as the tails of $x_1$.  \\

\noindent
\textbf{\large Case 4.3.1: $p_1$ is a P-node.} \\
  If $C_2-X = \emptyset$, then using the same argument as before (Lemma \ref{preliminaries}.0), we get $U(p_1) \subseteq U(n_2)$.
Hence we replace ${\cal L}_i$ and ${\cal L}'_i$ with ${\cal L}_i + {\cal L}'_i$ in ${\cal T}_1$ and ${\cal T}_2$
changing $U(n_1)$ to $U(n_1) \cap U(n_2) \supseteq U(p_1)$, and we match $n_1$ with $n_2$.

  Now we look at the case when $C_2-X$ is non-empty. Let $L=\brac{l_1,l_2, \cdots, l_{k_1}}$ be the set of
essential nodes appearing to the left of $X$ and $R = \brac{r_1, \cdots, r_{k_2}}$ be the set of essential nodes appearing to the right of $X$ in
${\cal T}_2$. Let $Y = \brac{y_1, \cdots, y_{k_3}}$ be all the remaining child nodes of $p_1$ other than $n_1$. For all
$i \in \brac{1,\cdots, k_3}$, if $MMD(y_i) \cap MMD(l_j) \not= \emptyset$ for some $j \in \brac{1,\cdots,k_1}$, then $y_i$ and $l_j$ both
have a common max-clique descendant say $Y$, and further in any leaf order of a common intersection tree ${\cal L}_1 + {\cal L}'_1$ must
appear between $Y$ and the descendants of $x_1$.

\begin{figure}
\begin{center}
\scalebox{0.6}{\includegraphics{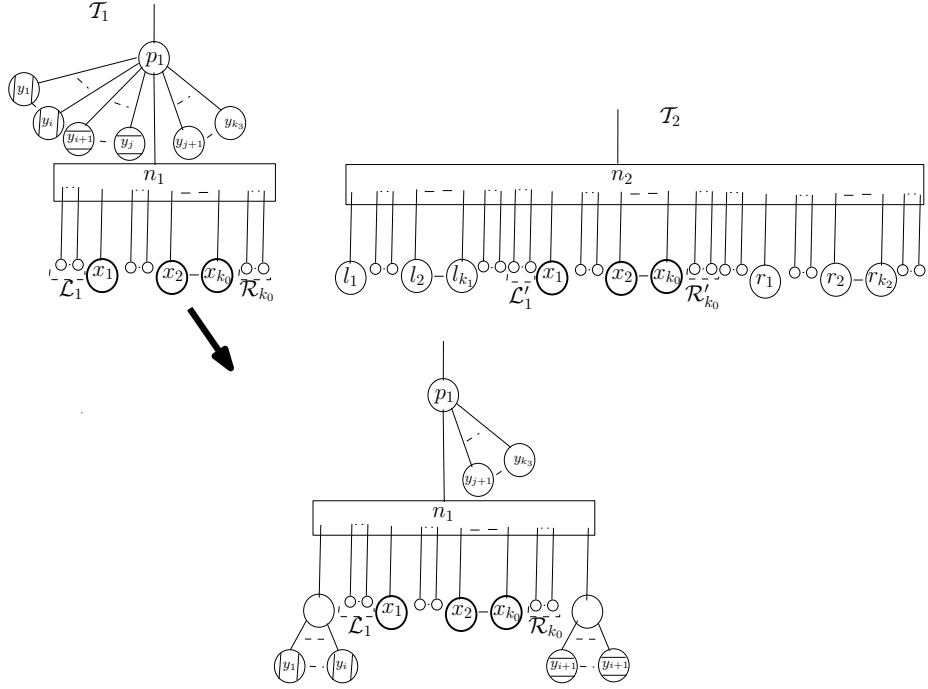}}
\end{center}
\caption{First reduction template of ${\cal T}_1$ for Case 4.3.1. A node $y_a$ has vertical stripes if $MMD(y_a) \cap MMD(l_b) \not= \emptyset$ for
some $l_b$, horizontal stripes if $MMD(y_a) \cap MMD(r_b) \not= \emptyset$ for some $r_b$ and no stripes otherwise.}
\label{qq4}
\end{figure}

  Thus we group all $y_i$ such that $MMD(y_i) \cap MMD(l_j) \not= \emptyset$ into a new P-node and add it to the (immediate) left of
${\cal L}_1$ (see Figure \ref{qq4}). Similarly, we group all $y_i$ such that $MMD(y_i) \cap MMD(r_j) \not= \emptyset$, for some
$j \in \brac{1, \cdots, k_2}$ into a new P-node and add it to the (immediate) right of ${\cal R}_{k_0}$.

  Note that if for some $y \in Y$, there exists $l_i$ and $r_j$ such that both $MMD(y) \cap MMD(l_i)$ and 
$MMD(y) \cap MMD(r_j) \not= \emptyset$, then we can conclude that ${\cal T}_1$ and ${\cal T}_2$ are incompatible.
 
  Also if $L$ and $R$ are both non-empty and  for all $y \in Y$, $MMD(y)$ doesn't intersect with any $MMD(l_i)$ for 
$i \in \brac{1, \cdots, k_1}$ and with any $MMD(r_j)$ for $j \in \brac{1,\cdots, k_2}$ then once again we conclude that ${\cal T}_1$
and ${\cal T}_2$ are incompatible.

\begin{figure}
\begin{center}
\scalebox{0.6}{\includegraphics{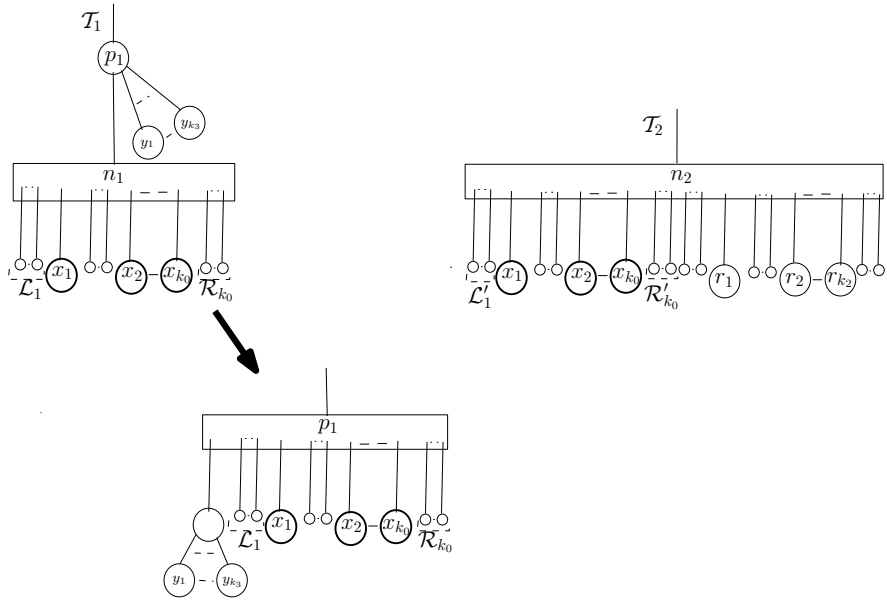}}
\end{center}
\caption{Second reduction template of ${\cal T}_1$ for Case 4.3.1. The stripes on the $y$ nodes are defined as before}
\label{qq5}
\end{figure}

  On the other hand if one of $L$ or $R$ is empty, say $L$, and $MMD(y_i) \cap MMD(r_j)$ is empty for all $y_i \in Y$ and $r_j \in R$, then the
above template would not reduce ${\cal T}_1$.  But then note that in any leaf-ordering of any intersection tree, ${\cal L}_1 + {\cal L}'_1$
should appear between the descendants of $y_i$ and $x_1$ for all $y_i \in Y$ (because of the constraints imposed by ${\cal T}_1$ and
${\cal T}_2$). Hence in this case we group all the nodes of $Y$ into a P-node and add it a child node of $p_1$ to the (immediate) left of
${\cal L}_1$ as shown in Figure \ref{qq5}.

  Note that since the MM-Descendents of any two sibling nodes are disjoint, both of the above templates can be implemented in $O(n)$ time. \\

\noindent
\textbf{\large Case 4.3.2: $p_1$ is a Q-node and $n_1$ is its only essential child.} \\
      Let ${\cal L}_p$ and ${\cal R}_p$ be the left and right tails of $p_1$. Note that in this case all the siblings of $n_1$ are subcliques that
are present in its tails. We have three subcases depending on how $U(p_1)$ intersects $U(n_2)$.

   Suppose $U(p_1)$ properly intersects $U(n_2)$.  We have $U(p_1)-U(n_2) \not= \emptyset$ and $U(n_2)-U(p_1) \not= \emptyset$.
We first claim that $C_2-X$ is empty. Suppose not. Let $Z$ be a max-clique descendant of a node in $C_2-X$ and $X_1$ be a max-clique
descendant of $x_1$. By Lemma \ref{preliminaries}.2, in ${\cal T}_1$, $Z$ appears outside the subtree rooted at $n_1$, and hence outside
the subtree rooted at $p_1$. Thus using Lemma \ref{preliminaries}.0, we conclude that each descendant of $p_1$ must contain all the vertices in
$Z \cap X_1 \supseteq U(n_2)$. A contradiction. Hence $C_2-X$ is empty.

 Now by Lemma \ref{preliminaries}.6,
there exists a subclique $S_1 \not\supseteq U(n_2)$  such that $S_1$ is the first clique of ${\cal L}_p$ or the last clique of ${\cal R}_p$.
Similarly there exists a subclique $S_2 \not\supseteq U(p_1)$ such that $S_1$ is the first clique of ${\cal L}'_1$ or the last clique of
${\cal R}'_{k_0}$. Without loss of generality let $S_1$ be the first clique of ${\cal L}_p$ and $S_2$ be the last clique of ${\cal R}'_{k_0}$.
Observe that
$S_1 \supseteq U(p_1)$ and $S_2 \supseteq U(n_2)$. This implies that $S_1$ and $S_2$ cannot be in the same tail (of $x_1$ or $x_{k_0}$) in
any intersection tree of ${\cal T}_1$ and ${\cal T}_2$. Thus we reduce ${\cal T}_1$ by collapsing $n_1$ i.e.~by deleting $n_1$, changing
the parent of child nodes of $n_1$ to $p_1$ and arranging the child nodes such that ${\cal L}_p$ appears to the left of ${\cal L}_1$ and
${\cal R}_p$ appears to the right of ${\cal R}_{k_0}$. Clearly, this reduction can be done in $O(n)$ time.

  Now we have to deal with the case when either $U(n_2) \subseteq U(p_1)$ or $U(p_1) \subseteq U(n_2)$.
Note that in any intersection tree ${\cal T}_I$ (of ${\cal T}_1$ and ${\cal T}_2$), the cliques of
${\cal L}_1 + {\cal L}'_1$ appear in the left tail of $x_1$ and the cliques of ${\cal R}_{k_0} + {\cal R}'_{k_0}$  appear in the right tail of
$x_{k_0}$. Further either (a) the cliques of ${\cal L}_p$ appear in the left tail of $x_1$ and the cliques of ${\cal R}_p$ appear in the right
tail of $x_{k_0}$ or (b) the cliques of ${\cal L}_p$ appear in the right tail of $x_{k_0}$ and the cliques of ${\cal R}_p$ appear in the left
tail of $x_1$. In the first case  ${\cal L}_1 + {\cal L}'_1 + {\cal L}_p$ and ${\cal R}_{k_0} + {\cal R}'_{k_0} + {\cal R}_p$ are both valid
and in the second case ${\cal L}_1 + {\cal L}'_1 + \bar{\cal R}_p$ and ${\cal R}_{k_0} + {\cal R}'_{k_0} + \bar{\cal L}_p$ are both valid. If
neither of these is valid then we conclude that ${\cal T}_1$ and ${\cal T}_2$ are incompatible. If exactly one of the above merges is valid,
then there is a unique way of collapsing $n_1$. When both of the above merge pairs are valid, we use the reduction template shown in
Figure \ref{qq2}. The justification (given below) depends on whether $U(n_2) \subseteq U(p_1)$ or $U(p_1) \subseteq U(n_2)$.

  Let $U(n_2) \subseteq U(p_1)$. Note that by Lemma \ref{preliminaries}.6, the intersection of the universal nodes of the first and last child
nodes of $p_1$ is $U(p_1)$. Hence if
${\cal L}_1 + {\cal L}'_1 + {\cal L}_p$, ${\cal R}_{k_0} + {\cal R}'_{k_0} + {\cal R}_p$, ${\cal L}_1 + {\cal L}'_1 + \bar{\cal R}_p$
and ${\cal R}_{k_0} + {\cal R}'_{k_0} + \bar{\cal L}_p$ are all valid then any subclique in ${\cal L}'_1$ or ${\cal R}'_{k_0}$ is either
a superset or a subset of $U(p_1)$. Thus in any intersection tree ${\cal T}_I$, any subclique $S$ in the left tail of $x_1$ or the
right tail of $x_{k_0}$ is either a superset or a subset of $U(p_1)$. Further if $S \supseteq U(p_1)$, then $S$ must appear in one of
$\brac{{\cal L}'_1, {\cal R}'_{k_0}, {\cal L}_p, {\cal R}_p, {\cal L}_1, {\cal R}_{k_0}}$. This implies that an intersection tree
satisfying condition (a) exists if and only if an intersection tree satisfying condition (b) exists. This justifies the use of our template
in Figure \ref{qq2}, for reducing ${\cal T}_1$.

  Similarly, if $U(p_1) \subseteq U(n_2)$, we infer that any clique in ${\cal L}_p$ or ${\cal R}_p$ is either a subset of $U(n_2)$ or a superset of
$U(n_2)$. This in turn implies that in ${\cal T}_I$, any subclique $S$ in the left tail of $x_1$ or the right tail of $x_1$, is
either a subset or a superset of $U(n_2)$. Further, if $S \supseteq U(n_2)$ then it must appear in
$\brac{{\cal L}_1, {\cal R}_{k_0}, {\cal L}'_1, {\cal R}'_{k_0}, {\cal L}_p, {\cal R}_p}$. This implies that an intersection tree
satisfying condition (a) exists if and only if an intersection tree satisfying (b) exists. This justifies the use of our template
in Figure \ref{qq2}, for reducing ${\cal T}_1$. 

  We now show that the amortized cost of executing the reduction template in Figure \ref{qq2}, over all instances of the algorithm takes $O(n^2)$ time.
Note that we use the same template for Case 4.3.3 when $C_2-X$ is empty. It is enough to show that the amortized time of all the mergability checks:
(whether ${\cal L}'_1 + {\cal L}_p$ and ${\cal R}'_{k_0} + {\cal R}_p$ are both valid) take $O(n^2)$ time.

   Let $c({\cal L}'_1)$ and $c({\cal R}'_{k_0})$ be the (consecutive) subsequences of ${\cal L}'_1$ and ${\cal R}'_{k_0}$ (respectively) such that each subclique in $c({\cal L}'_1)$ and $c({\cal R}'_{k_0})$ contains $U(p_1)$ but not $U(n_1)$. $c({\cal L}'_1)$ and $c({\cal R}'_{k_0})$ are said to be
the core tails of $n_2$.

   Similarly let $c({\cal L}_p)$ and $c({\cal R}_p)$ be the (consecutive) subsequences of ${\cal L}_p$ and ${\cal R}_p$ (respectively) such that each
subclique in $c({\cal L}_p)$ and $c({\cal R}_p)$ contains $U(n_2)$. $c({\cal L}_p)$ and $c({\cal R}_p)$ are said to be the {\it core} tails of $p_1$.

   Note that the core tails are only defined for $p_1$ and $n_2$, for the current case and Case 4.3.3, when $C_2-X$ is empty. We define the core
tails of all other nodes to be empty. Observe that when $n_1$ is collapsed, the (new) core tails of any node in ${\cal T}_1$
(resp. ${\cal T}_2$) are disjoint from the core tails of $p_1$ (resp. $n_2$) before the collapse.
 
   We observe that checking the validity of ${\cal L}'_1 +{\cal L}_p$ reduces to checking the validity of
$c({\cal L}'_1) +c({\cal L}_p)$. Similarly, checking the validity of ${\cal R}'_{k_0} + {\cal R}_p$ reduces to checking the validity of
$c({\cal R}'_{k_0}) + c({\cal R}_p)$. 

  Now computing the cores over all executions of this template , takes $O(n \cdot |I|) = O(n^2)$ amortized time. Also,
computing the mergability of the cores, over all executions of the template takes ${\sum_i (m_i + t_i) |I|}$, where $m_i$, $t_i$
 are the number of subcliques in the cores of $n_2$ and $p_1$ (respectively), in the $i$th execution of the template. Since
${\sum_i m_i} = O(n)$ and ${\sum_i t_i} = O(n)$, the total running time of template 10 over all executions is $O(n^2)$. \\

\begin{figure}
\begin{center}
\scalebox{0.6}{\includegraphics{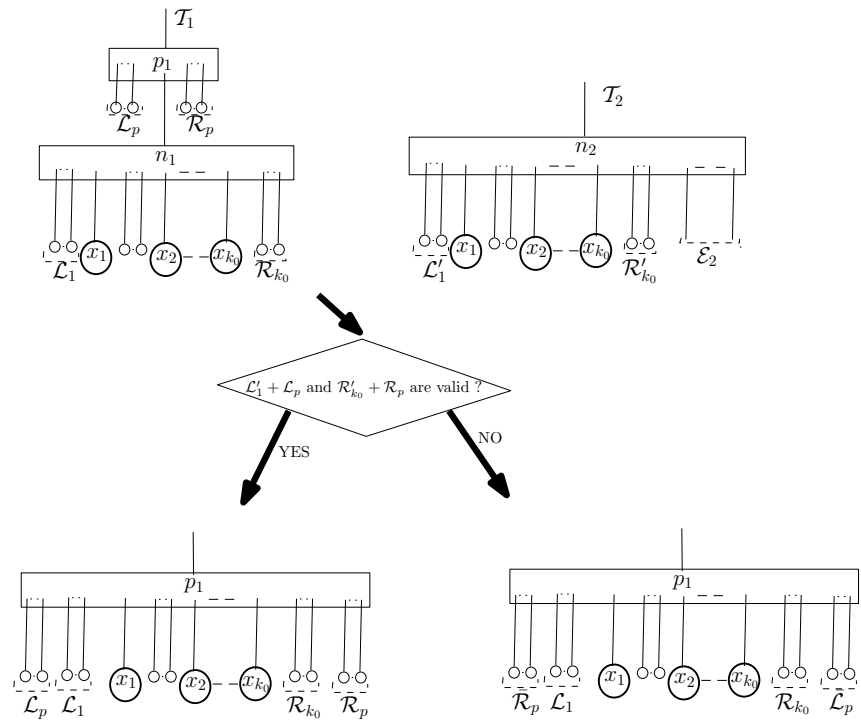}}
\end{center}
\caption{Reduction template of ${\cal T}_1$ for Case 4.3.2. Note that ${\cal E}_2$ denotes the (possibly empty) sequence of children of $n_2$
that appear after ${\cal R}'_{k_0}$.}
\label{qq2}
\end{figure}

\noindent
\textbf{\large Case 4.3.3: $p_1$ is a Q-node with more than one essential child.} \\
  Now let $y$ be an essential child of $p_1$, such that all the nodes between $n_1$ and $y$ are subcliques. Without loss of generality,
we assume that $y$ appears to the right of $n_1$.

  We first consider the subcase when $C_2-X$ is empty. In this case observe that all the max-clique descendants of $y$ appear outside the
subtree rooted at $n_1$ in ${\cal T}_1$.
Applying Lemma \ref{preliminaries}.0 on a max-clique descendant of $y$ and a max-clique descendant of $x_1$, we infer that each descendant
clique of $n_2$ must contain all the vertices in $U(p_1)$. In other words we get $U(p_1) \subseteq U(n_2)$. Now
the template (and the argument) in this case is analogous to case 4.3.2, when $U(p_1) \subseteq U(n_2)$ (Figure \ref{qq2}).

\begin{figure}
\begin{center}
\scalebox{0.6}{\includegraphics{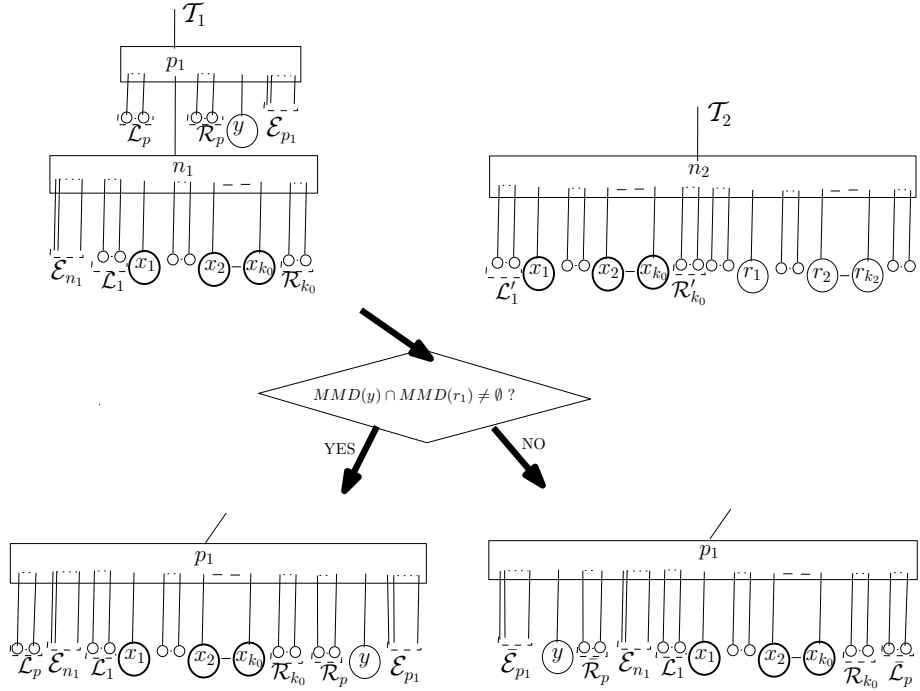}}
\end{center}
\caption{Reduction template of ${\cal T}_1$ for Case 4.3.3.}
\label{qq3}
\end{figure}

 Now suppose $C_2-X$ is non empty. Let $r_1$ be
the first essential child to the right of $x_{k_0}$ in ${\cal T}_2$.
We use the templates in Figure \ref{qq3}, depending on whether $MMD(y) \cap MMD(r_1)$ is empty
or not. If $MMD(y) \cap MMD(r_1)$ is non-empty, then by Lemma \ref{preliminaries}.2, there exists a max-clique $Y$ that is a descendant of both
$r_1$ and $y$.
Now if ${\cal T}_1$ and ${\cal T}_2$ are compatible, then in any leaf ordering of an intersection tree
${\cal T}_I$, the subcliques of ${\cal R}_{k_0} + {\cal R}'_{k_0}$ appear in between the descendants
of $x_k$ and $Y$ (because of ${\cal T}_2$). This justifies the reduction template of ${\cal T}_1$ in Figure \ref{qq3}.

  On the other hand, if $MMD(r_1) \cap MMD(y) = \emptyset$, then by the constraints of ${\cal T}_2$, in any leaf ordering of ${\cal T}_I$,
the subcliques of ${\cal R}_{k_0} + {\cal R}'_{k_0}$ appear between the descendants of $x_{k_0}$ and $l_1$ and further no max-clique appears
between them. This justifies the reduction template of ${\cal T}_1$ in Figure \ref{qq3}. Moreover the template reduction takes $O(n)$ time.\\

\noindent
\textbf{Run time of the Algorithm}\\
 In this Section, we show that the run time of our algorithm is $O(n^2{\log n})$, where $n$ is the total number of vertices in $G_1 \cup G_2$. 

 Observe that reducing ${\cal T}_1$ [resp. ${\cal T}_2$] decreases the number of leaf orderings of ${\cal T}_1$ [resp. ${\cal T}_2$] by at least 
half. Moreover the total number of nodes in ${\cal T}_1$ and ${\cal T}_2$ is at most $n$. Thus the number of leaf orderings of ${\cal T}_1$ and 
${\cal T}_2$ is at most $n!$ and hence the algorithm requires at most $n {\log n}$ reductions.

We begin by showing that selecting the nodes $n_1$ and $n_2$ takes $O(n)$ time in any iteration. We first note that computing the number of 
MM-Descendants for all the nodes takes $O(n)$ time (they can be computed in a bottom-up fashion). With each node $x$, we store 
$U(x)$ and the cardinality of $MMD(x)$.

Recall that as we go down the tree the universal sets increase and the MM-Descendants sets decrease. Thus $n_1$ must have
maximal depth among all unmatched essential nodes. Hence we can select $n_1$ in $O(n)$ time by looking at the unmatched essential
nodes of maximal depth in ${\cal T}_1$ and ${\cal T}_2$, and selecting a node with the greatest universal set size and the least number 
of MM-Descendants in that order. Note that by property 1 of Lemma \ref{properties}, the MM-descendants of $n_1$ are same as the essential
child nodes of $n_1$. Now we can select $n_2$ from the other tree ${\cal T}_2$ in $O(n)$ time as follows: Let $S$ be the set of [matched]
essential children of $n_1$ and $S'$ be the corresponding set of matched nodes in ${\cal T}_2$. Let $p(S')$ be the set of parent nodes
of nodes in $S'$. We select $n_2$ to be a node of maximum depth among $p(S')$.

Also recall that at the high-level our algorithm has 4 cases depending on whether $n_1$, $n_2$ are P-nodes or Q-nodes. We showed that each step in
cases 1,2 or 3 takes  $O(n)$ time. For case 4, we showed that each of reduction steps, excluding the mergability checks in Cases 4.3.2 and 4.3.3
take $O(n)$ time. We also showed that the mergability checks of Cases 4.3.2 and 4.3.3 take $O(n^2)$ amortized time over all steps of the algorithm.
Further, in the beginning of Case 4, we showed that the matching steps, which involve inserting the subcliques of one tree into the other take at most
$O(n^2)$ amortized time.  Thus the total time taken by our algorithm is $O(n^2 {\log n} + n^2 + n^2) = O(n^2{\log n})$.
At each node $y$ of ${\cal T}_1$ (resp. ${\cal T}_2$) we explicitly store the set $U(y)$ and the cardinality of $MMD(y)$. Since the number of internal 
nodes is less than the number of leaf nodes, this additional storage still takes $O(n+m)$. Thus the space complexity of our algorithm is
$O(n+m)$.

\section{Open Problem}
  Simultaneous graphs can be generalized in a natural way to more than two graphs: when
$G_1 = (V_1, E_1), G_2 = (V_2, E_2), \cdots, G_k = (V_k, E_k)$ are $k$ graphs in class $\cal C$,
sharing a vertex set $I$ and its induced edges i.e.~$V_i\cap V_j = I$ for all $i,j \in \brac{1,\cdots, k}$.
In this version of the problem the set of optional edges induces a complete $k$-partite graph and hence
this also generalizes probe graphs. This generalized version can be solved in polynomial time for comparability 
and permutation graphs~\cite{JL2}. We conjecture that it can be solved in polynomial time for interval graphs.

\bibliography{newref.bib}